\documentclass[10pt,aps,prx,twocolumn,amsfonts,floatfix]{revtex4}
\usepackage[ascii]{inputenc}
\usepackage{amsfonts}
\usepackage{amsmath}
\usepackage{amssymb}
\usepackage{amsthm}
\usepackage{tikz}
\usetikzlibrary{decorations.pathreplacing,decorations.pathmorphing,calc}
\usepackage{graphicx}
\usepackage[caption=false]{subfig}
\usepackage{bm}
\usepackage{nicefrac}       
\usepackage{braket}
\usepackage{mathtools}
\usepackage{color,colortbl}
\usepackage{hyperref}
\usepackage{algorithm}
\usepackage{algpseudocode}

\DeclareMathOperator{\e}{e}
\DeclareMathOperator{\tr}{tr}

\DeclareMathOperator*{\argmin}{argmin}

\newcommand{\N}{\mathbb{N}}

\newcommand{\R}{\mathbb{R}}
\newcommand{\C}{\mathbb{C}}

\newcommand{\abs}[1]{\lvert{#1}\rvert}
\newcommand{\norm}[1]{\lVert{#1}\rVert}

\newtheorem{theorem}{Theorem}

\graphicspath{{figures/}}

\begin{document}

\title{Quantum Process Tomography of Unitary Maps from Time-Delayed Measurements}

\author{Irene L\'opez Guti\'errez}
\email{irene.lopez@tum.de}
\affiliation{Technical University of Munich, Department of Informatics, Boltzmannstra{\ss}e 3, 85748 Garching, Germany}

\author{Felix Dietrich}
\email{felix.dietrich@tum.de}
\affiliation{Technical University of Munich, Department of Informatics, Boltzmannstra{\ss}e 3, 85748 Garching, Germany}

\author{Christian B.~Mendl}
\email{christian.mendl@tum.de}
\affiliation{Technical University of Munich, Department of Informatics, Boltzmannstra{\ss}e 3, 85748 Garching, Germany}
\affiliation{Technical University of Munich, Institute for Advanced Study, Lichtenbergstra{\ss}e 2a, 85748 Garching, Germany}

\begin{abstract}
Quantum process tomography conventionally uses a multitude of initial quantum states and then performs state tomography on the process output. Here we propose and study an alternative approach which requires only a single (or few) known initial states together with time-delayed measurements for reconstructing the unitary map and corresponding Hamiltonian of the time dynamics. The overarching mathematical framework and feasibility guarantee of our method is provided by the Takens embedding theorem. We explain in detail how the reconstruction of a single qubit Hamiltonian works in this setting, and provide numerical methods and experiments for general few-qubit and lattice systems with local interactions. In particular, the method allows to find the Hamiltonian of a two qubit system by observing only one of the qubits.
\end{abstract}

\maketitle

\section{Introduction}

System identification refers to the estimation of the dynamics of a system from measurements of its characteristics. Its quantum analogue, quantum process tomography (QPT), is essential for the realization and testing of quantum devices \cite{childs-2001, obrien-2004, bialczak-2010}, as a benchmarking tool for quantum algorithms \cite{weinstein-2004, kampermann-2005}, and in general for understanding the inner workings of a quantum system \cite{jullien-2014, bisognin-2019}. However, textbook algorithms for QPT \cite{nielsen-2010} might be difficult to realize in laboratory settings due to the required preparation of many initial states and observation of the complete target system.

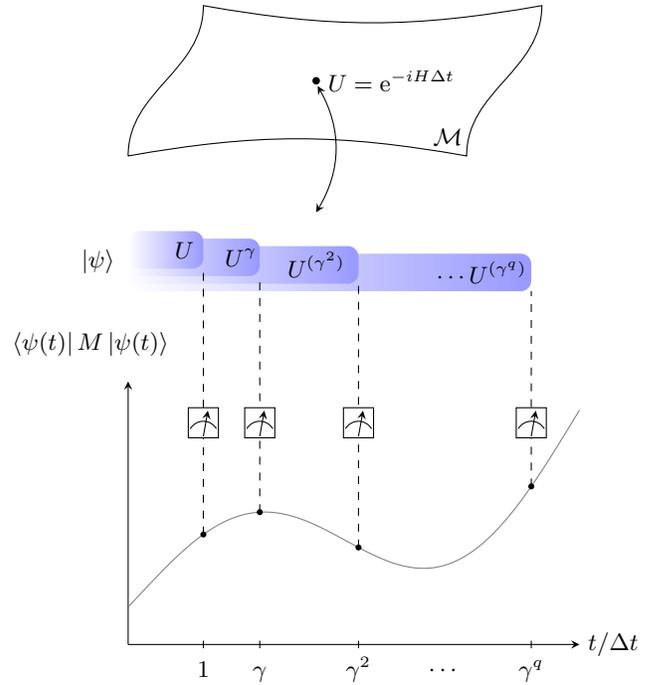
\begin{figure}
\centering
\begin{tikzpicture}[>=stealth]
\begin{scope}[shift={(0,6.5)}]
\draw (0,0) to[out=10, in=170] (4.5,0) to[out=90,in=-90] (5.5,2) to[out=190, in=-10] (1,2) to[out=-90, in=90] (0,0);
\coordinate (U) at (2.5,1);
\fill (U) circle (0.05);
\node at (3.5,1) {$U = \e^{-i H \Delta t}$};
\node at (4.25,0.25) {$\mathcal{M}$};
\draw[<->] ($(U)+(0.02,-0.07)$) to[bend left] (2.5,-0.75);
\end{scope}
\begin{scope}[shift={(0,5)}]
\node (psi) at (-0.4,0.1) {$\ket{\psi}$};
\shade[right color=blue!40, left color=white, rounded corners] (0,-0.3) rectangle (343/64,0.2);
\shade[right color=blue!40, left color=white, rounded corners] (0,-0.2) rectangle (49/16,0.3);
\shade[right color=blue!40, left color=white, rounded corners] (0,-0.1) rectangle (7/4,0.4);
\shade[right color=blue!40, left color=white, rounded corners] (0, 0)   rectangle (1,0.5);
\node at (0.75,0.25) {$U$};
\node at (1.5, 0.15) {$U^{\gamma}$};
\node at (2.5, 0.05) {$U^{(\gamma^2)}$};
\node at (4.7,-0.05) {$\cdots U^{(\gamma^q)}$};
\end{scope}
\draw[dashed] (1,     1.46391) -- (1,5);
\draw[dashed] (7/4,   1.76065) -- (7/4,4.9);
\draw[dashed] (49/16, 1.29102) -- (49/16,4.8);
\draw[dashed] (343/64,2.10366) -- (343/64,4.7);
\foreach \t in {1,7/4,49/16,343/64}
{
    \begin{scope}[shift={(\t-0.2,2.75)}, scale={0.4}]
    \draw[fill=white] (0,0) rectangle (1,1);
    \draw (0.5, 0.55) arc (90:150:0.5);
    \draw (0.5, 0.55) arc (90: 30:0.5);
    \draw[->] (0.5, 0.05) -- +(80:0.8660254166);
    \end{scope}
}
\draw[->] (0,0) -- (6,0) node[right] {$t / \Delta t$};
\draw[->] (0,0) -- (0,3.5);
\node at (-0.5, 4) {$\bra{\psi(t)} M \ket{\psi(t)}$};
\draw[thin] (1,0.05)      -- (1,-0.05)      node[below] {$1\vphantom{\gamma^2}$};
\draw[thin] (7/4,0.05)    -- (7/4,-0.05)    node[below] {$\gamma\vphantom{\gamma^2}$};
\draw[thin] (49/16,0.05)  -- (49/16,-0.05)  node[below] {$\gamma^2$};
\draw[thin] (343/64,0.05) -- (343/64,-0.05) node[below] {$\gamma^q\vphantom{\gamma^2}$};
\node at (539/128,-0.35) {$\cdots$};
\draw[gray] plot [smooth] coordinates {(0., 0.5) (0.1, 0.605515) (0.2, 0.711858) (0.3, 0.817893) (0.4, 0.922507) (0.5, 1.02462) (0.6, 1.12321) (0.7, 1.2173) (0.8, 1.30598) (0.9, 1.38843) (1., 1.46391) (1.1, 1.53178) (1.2, 1.5915) (1.3, 1.64264) (1.4, 1.68487) (1.5, 1.71799) (1.6, 1.74191) (1.7, 1.75666) (1.8, 1.7624) (1.9, 1.75938) (2., 1.74799) (2.1, 1.72871) (2.2, 1.70212) (2.3, 1.66891) (2.4, 1.62985) (2.5, 1.58579) (2.6, 1.53764) (2.7, 1.48639) (2.8, 1.43306) (2.9, 1.37871) (3., 1.32443) (3.1, 1.2713) (3.2, 1.22043) (3.3, 1.1729) (3.4, 1.12976) (3.5, 1.09202) (3.6, 1.06066) (3.7, 1.03657) (3.8, 1.02061) (3.9, 1.01351) (4., 1.01594) (4.1, 1.02847) (4.2, 1.05156) (4.3, 1.08557) (4.4, 1.13072) (4.5, 1.18713) (4.6, 1.25481) (4.7, 1.33364) (4.8, 1.42336) (4.9, 1.52363) (5., 1.63397) (5.1, 1.7538) (5.2, 1.88244) (5.3, 2.01909) (5.4, 2.16288) (5.5, 2.31287) (5.6, 2.46803) (5.7, 2.62727) (5.8, 2.78947) (5.9, 2.95345) (6., 3.11803)};
\fill (1,     1.46391) circle (0.04);
\fill (7/4,   1.76065) circle (0.04);
\fill (49/16, 1.29102) circle (0.04);
\fill (343/64,2.10366) circle (0.04);
\end{tikzpicture}
\caption{Schematic illustration of the map from a unitary time step operator $U$ to a vector of measurement averages at different time delays.}
\label{fig:schematic}
\end{figure}

In this work we focus on the unitary time evolution of a closed quantum system. We propose and investigate an approach based on measurements with different \emph{time delays}, which should be easily realizable in lab experiments. Our main contribution are algorithms and numerical procedures for identifying the corresponding time evolution operator, and thus indirectly the quantum Hamiltonian. Intriguingly, we can identify the operator for the entire system even if the measurements are restricted to a subsystem (using two known initial quantum states). The Takens embedding theorem, discussed in Sect.~\ref{sec:takens}, provides an overarching mathematical foundation and feasibility guarantee for our approach. Concretely, the mathematical framework starts with a manifold $\mathcal{M}$, which we take to be the special unitary group of time evolution operators. Given an initial quantum state, a time step matrix $U \in \mathcal{M}$ then determines the measurement averages at a sequence of time points, see Fig.~\ref{fig:schematic}. The Takens theorem states that the map from $\mathcal{M}$ to this measurement vector is actually a smooth embedding (under certain conditions), which then allows us to reconstruct $U$.

Related to the present work, a recent approach for quantum process tomography using tensor networks as a parametrization of the quantum channel was able to achieve high accuracies for systems of up to 10 qubits \cite{torlai-2020}. Furthermore, in \cite{baldwin-2014} the authors derive a lower bound in the number of POVMs required to fully characterize a unitary or near-unitary map. Regarding the related task of quantum state tomography, several methods, some based on machine learning techniques, have been developed \cite{huang-2020, cramer-2010, torlai-2018, flurin-2020}. However, except for \cite{flurin-2020}, these works do not explicitly take advantage of the time trajectory of the quantum system as envisioned here. Gate Set Tomography~\cite{Nielsen-2021} does not need pre-calibrated quantum states, but relies on very long time series.

Our work is closely related to earlier research on the identification of quantum Hamiltonians from time series~\cite{Zhang-2014c,Zhang-2015}. The authors obtain all degrees of freedom of the Hamiltonian with known structure by solving a system of equations, mostly in the presence of noise. More recently,  neural networks have been used for identification, even from experimental data~\cite{Xin-2019,Che-2021,Zhao-2021}. The minimal number of observables needed to identify the Hamiltonian is not addressed, and we provide it here with the link to Takens theorem. In our work, the choice of initial state is not as important, as we will demonstrate. 

Takens theorem has been fundamental to several system identification methods for general dynamical systems already~\cite{sauer-1994}, recently also involving machine learning~\cite{berry-2013,dietrich-2016b,dietrich-2020,Giannakis-2021a}, with a focus on PDE~\cite{kemeth-2018,Kemeth-2020} or special structure such as (classical) Hamiltonian dynamics~\cite{bertalan-2019,Greydanus-2019}.
The identification of a unitary map from measurement data has been discussed by Koopman and von Neumann~\cite{koopman-1931,koopman-1932}, work that has been revived in a data-driven context in the last twenty years and extended to dissipative systems~\cite{mezic-2005,Mezic-2019c,dietrich-2020b}.
The connection to quantum systems and their special structure and challenges has not yet been addressed in the work cited above, however.

\section{Physical model}
\label{sec:model}

\begin{figure}[!ht]
\centering
\includegraphics[width=\columnwidth]{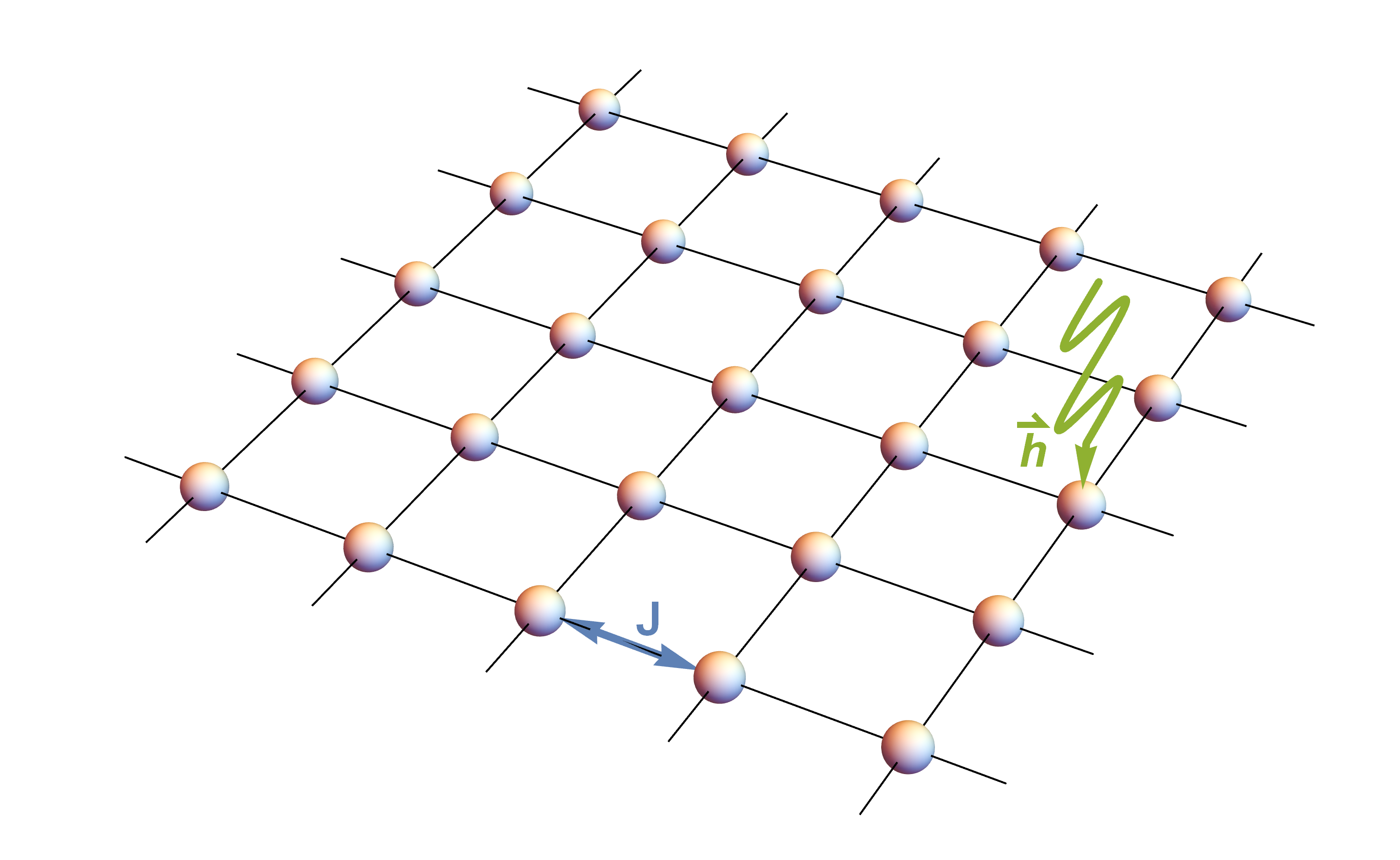}
\caption{Visualization of the quantum Hamiltonian in Eq.~\eqref{eq:HIsing} on a two-dimensional lattice, with $J$ the interaction strength and $\vec{h}$ the external field.}
\label{fig:lattice_hamiltonian}
\end{figure}

We assume throughout that the quantum Hamiltonian $H$ is time-independent and will investigate two settings: (i) a few-qubit system and $H$ a general dense Hermitian matrix, and (ii) a (tight binding) Ising-type model on a two-dimensional lattice $\Lambda$, as widely studied in condensed matter physics. For concreteness, the physical system for case (ii) consists a local spin degree of freedom at each lattice site. The Hamiltonian is then defined as
\begin{equation}
\label{eq:HIsing}
H = - \sum_{\langle j, \ell \rangle} J_{j,\ell} \, \sigma^z_j \sigma^z_\ell - \sum_{j \in \Lambda} \vec{h}_j \cdot \vec{\sigma}_j,
\end{equation}
where $J_{j,\ell} \in \mathbb{R}$ and $\vec{h}_j \in \mathbb{R}^3$ are local parameters defining the interaction strength and the external field, respectively. $\sigma^{\alpha}_j$ for $\alpha \in \{x, y, z\}$ is the $\alpha$-th Pauli matrix acting on site $j \in \Lambda$, and $\vec{\sigma}_j = (\sigma^x_j, \sigma^y_j, \sigma^z_j)$ the corresponding Pauli vector. The first sum in \eqref{eq:HIsing} runs over nearest neighbors on the lattice. We take $\Lambda$ to contain a finite number $n$ of sites, and assume periodic boundary conditions. The site-dependent parameters allow to simulate disorder. Fig.~\ref{fig:lattice_hamiltonian} illustrates the interaction and external field terms of $H$ on a two-dimensional lattice. The precise form of $H$ is not important for the reconstruction as long as the time dynamics it generates can be well approximated by a quantum circuit, for example via Trotterization. We assume that the overall structure of $H$ is known, and our task is to determine the numerical values of the parameters.

For later reference, we state the unitary time evolution operator at time $t \in \R$ based on the Schr\"odinger equation (in units of $\hbar = 1$):
\begin{equation}
\label{eq:time_evolve_unitary}
U(t) = \e^{-i H t}.
\end{equation}
The solution to the Schr\"odinger equation for an initial wavefunction (statevector) $\psi \in \C^N$ is then $\psi(t) = U(t) \psi$.

\section{Takens embedding framework}
\label{sec:takens}

The theorems of Takens and Ruelle~\cite{ruelle-1971,takens-1981}, based on the embedding theorems of Whitney~\cite{whitney-1936}, are the theoretical foundations of time-delay embedding we use here.

The general idea we employ in this paper is to ``embed'' the manifold of unitary matrices (describing the dynamics of a quantum system) into the space of measurement trajectories. We first state the mathematical theorem, and then discuss the specialization for quantum time evolution. Let $k \geq d \in \N$, and $\mathcal{M} \subset \R^k$ be a $d$-dimensional, compact, smooth, connected, oriented manifold with Riemannian metric $g$ induced by the embedding in $k$-dimensional Euclidean space. Note that this setting is sufficient for our presentation, but is more restrictive than allowed by the results cited below.

Together with the results from Packard et al.~\cite{packard-1980} and Aeyels~\cite{aeyels-1981}, the definitions and theorems of Takens~\cite{takens-1981} describe the concept of observability of state spaces of nonlinear dynamical systems. A dynamical system is defined through its state space (here, the manifold $\mathcal{M}$) and a diffeomorphism $\phi: \mathcal{M} \to \mathcal{M}$.
\begin{theorem}\textbf{Generic delay embeddings}\label{thm:takens_1}
For pairs $(\phi,y)$, $\phi:\mathcal{M} \to \mathcal{M}$ a smooth diffeomorphism and $y: \mathcal{M} \to \mathbb{R}$ a smooth function, it is a generic property that the map $\Phi_{(\phi,y)}: \mathcal{M} \to \mathbb{R}^{2d+1}$, defined by
\begin{equation}
\label{eq:takens_embedding_map}
\Phi_{(\phi,y)}(x) = \Big( y(x), y(\phi(x)), \dots, y(\underbrace{\phi\circ\dots\circ\phi}_{2d~\text{times}}(x)) \Big)
\end{equation}
is an embedding of $\mathcal{M}$; here, ``smooth'' means at least $C^2$.
\end{theorem}
Genericity in this context is defined as ``an open and dense set of pairs $(\phi,y)$'' in the $C^2$ function space. Open and dense sets can have measure zero, so Sauer et al.~\cite{sauer-1991} later refined this result significantly by introducing the concept of prevalence (a ``probability one'' analog in infinite dimensional spaces). See \cite{stark-1997} for similar results with stochastic systems.

Let $N$ denote the quantum Hilbert space dimension. In our context, $\mathcal{M}$ is the special unitary group $\mathrm{SU}(N) \subset \C^{N \times N}$ when identifying $\C \simeq \R^2$. In particular, $\mathcal{M}$ (with underlying field $\R$) has dimension $d = N^2 - 1$. In physical terms, the elements of $\mathrm{SU}(N)$ are the unitary time evolution matrices in Eq.~\eqref{eq:time_evolve_unitary}. We assume that the Hamiltonian $H$ is traceless, since adding multiples of the identity to $H$ leads to a global phase factor in the time evolution, which is unobservable in subsequent measurements as envisioned here. We fix a time step $\Delta t$, which we may (without loss of generality) absorb into $H$, and set $U = \e^{-i H}$ in the following.

Now define the diffeomorphism $\phi$ via a scaling of $H$ by a factor $\gamma > 0$, $\gamma \neq 1$, such that
\begin{equation}
\label{eq:phi_gamma}
\phi(U) = \e^{-i \gamma H} = U^\gamma.
\end{equation}

Regarding $y$, fix a randomly chosen initial quantum state $\psi \in \C^{N}$ and an observable (Hermitian matrix) $M$. Now let $y$ compute the corresponding expectation value:
\begin{equation}
y: \mathcal{M} \to \mathbb{R}, \quad y(U) = \bra{\psi} U^{\dagger} M U \ket{\psi}.
\end{equation}
With these definitions, the output of the map $\Phi_{(\phi,y)}$ in Eq.~\eqref{eq:takens_embedding_map} becomes
\begin{multline}
\Phi_{(\phi,y)}(U) = \Big( \bra{\psi} \e^{i H} M \e^{-i H} \ket{\psi}, \\ \bra{\psi} \e^{i \gamma H} M \e^{-i \gamma H} \ket{\psi}, \\ \dots, \bra{\psi} \e^{i \gamma^{2d} H} M \e^{-i \gamma^{2d} H} \ket{\psi} \Big),
\end{multline}
physically corresponding to measurements at time points $t_q = \gamma^q$ for $q = 0, \dots, 2d$.

The main point here is the prospect to identify the time step matrix $U$, and thus indirectly the Hamiltonian, based on a single measurement time trajectory, under the assumption that the initial quantum state is known. As caveat, the relation $U = \e^{-i H \Delta t}$ determines the eigenvalues of $H$ only up to multiples of $2 \pi / \Delta t$. Moreover, the map $\Phi_{(\phi,y)}$ might not be one-to-one, in the sense that two different unitary matrices give rise to the same measurement trajectory. We discuss such a case in more detail in the following. Such issues can be avoided in practice by additional assumptions on the structure of $H$.

\section{Numerical methods}
\label{sec:methods}

Throughout this work, we assume that it is feasible to reliably prepare a single (or when indicated several) known initial state(s) $\psi \in \C^{N}$. To demonstrate the general applicability of our methods, $\psi$ is chosen at random in the following algorithms and numerical simulations. Specifically, the entries of $\psi$ before normalization are independent and identically distributed (i.i.d.) random numbers sampled from the standard complex normal distribution.

\subsection{Reconstruction algorithm for a single-qubit system}
\label{sec:bloch_single_qubit}

\begin{figure}[!ht]
\centering
\subfloat[trajectory on Bloch sphere]{%
\begin{tikzpicture}[>=stealth, scale=2]
\draw[thick] (0, 0) circle (1);
\pgfsetxvec{\pgfpoint{-0.309017cm}{-0.363954cm}}
\pgfsetyvec{\pgfpoint{ 0.951057cm}{-0.118256cm}}
\pgfsetzvec{\pgfpoint{ 0cm       }{ 0.92388cm}}
\draw[->] (0, 0, 0) -- (1.35, 0, 0) node[left]  {$x$};
\draw[->] (0, 0, 0) -- (0, 1.35, 0) node[right] {$y$};
\draw[->] (0, 0, 0) -- (0, 0, 1.35) node[left]  {$z$};
\draw[dashed] (0, 0) circle (1);
\draw[->] (0, 0, 0) -- (0.569803,0.683763,0.455842);
\node[fill=white, inner sep=0.5] at (0.7, 0.8, 0.6) {$\vec{h}$};
\draw[->] (0, 0, 0) -- (0.815958,0.355557,0.455842) ;
\node at (0.7, 0.2, 0.2) {$-\vec{h}'$};
\draw[->] (0, 0, 0) -- (0.565685,0.424264,0.707107);
\node at (0.6, 0.5, 0.85) {$\vec{r}$};
\draw[->, thick]
(0.565685,0.424264,0.707107) -- (0.610659,0.404219,0.680957) -- (0.653712,0.389959,0.648531) -- (0.693783,0.381835,0.610628) -- (0.729886,0.380047,0.568182) -- (0.761132,0.384638,0.522237) -- (0.786751,0.395497,0.473926) -- (0.806112,0.412356,0.424436) -- (0.81874,0.434798,0.374987) -- (0.824322,0.462273,0.326797) -- (0.822722,0.494103,0.281053) -- (0.813979,0.529504,0.238879) -- (0.798308,0.567606,0.201316) -- (0.776095,0.607469,0.169288) -- (0.747887,0.648112,0.143583) -- (0.714378,0.688535,0.124835) -- (0.676394,0.727741,0.113505) -- (0.634871,0.764766,0.109872) -- (0.590829,0.798698,0.114026) -- (0.545354,0.828702,0.125864) -- (0.499566,0.854038,0.145095) -- (0.454592,0.874083,0.171245) -- (0.41154,0.888343,0.20367) -- (0.371469,0.896467,0.241573) -- (0.335366,0.898255,0.284019) -- (0.30412,0.893664,0.329964) -- (0.278501,0.882805,0.378276) -- (0.25914,0.865947,0.427765) -- (0.246512,0.843504,0.477214) -- (0.24093,0.816029,0.525404) -- (0.24253,0.784199,0.571149);
\draw[dashed, red] (0.824334, 0.47926577, 0.30129351) -- (0,0,0.30129351);
\draw[red, fill=red] (0.824334, 0.47926577, 0.30129351) circle (0.02);
\draw[red, fill=red] (0,0,0.30129351) circle (0.02);
\draw[dashed, red] (0.78987563, 0.58395424, 0.18733378) -- (0,0,0.18733378);
\draw[red, fill=red] (0.78987563, 0.58395424, 0.18733378) circle (0.02);
\draw[red, fill=red] (0,0,0.18733378) circle (0.02);
\draw[dashed, red] (0.67083989, 0.73301341, 0.11253969) -- (0,0,0.11253969);
\draw[red, fill=red] (0.67083989, 0.73301341, 0.11253969) circle (0.02);
\draw[red, fill=red] (0,0,0.11253969) circle (0.02);
\draw[dashed, red] (0.45548524, 0.87373396, 0.17065218) -- (0,0,0.17065218);
\draw[red, fill=red] (0.45548524, 0.87373396, 0.17065218) circle (0.02);
\draw[red, fill=red] (0,0,0.17065218) circle (0.02);
\draw[dashed, red] (0.25499714, 0.86033713, 0.44135754) -- (0,0,0.44135754);
\draw[red, fill=red] (0.25499714, 0.86033713, 0.44135754) circle (0.02);
\draw[red, fill=red] (0,0,0.44135754) circle (0.02);

\draw[->, thick]
(0.565685,0.424264,0.707107) -- (0.559035,0.473052,0.680957) -- (0.5574,0.518375,0.648531) -- (0.560821,0.559118,0.610628) -- (0.569213,0.594277,0.568182) -- (0.58237,0.622988,0.522237) -- (0.599968,0.644541,0.473926) -- (0.621573,0.658408,0.424436) -- (0.646654,0.664247,0.374987) -- (0.674592,0.661913,0.326797) -- (0.704701,0.651464,0.281053) -- (0.736238,0.633159,0.238879) -- (0.768428,0.607446,0.201316) -- (0.800477,0.57496,0.169288) -- (0.831596,0.5365,0.143583) -- (0.861019,0.493013,0.124835) -- (0.888022,0.445571,0.113505) -- (0.911939,0.395341,0.109872) -- (0.932183,0.34356,0.114026) -- (0.948253,0.291504,0.125864) -- (0.959755,0.240453,0.145095) -- (0.966406,0.191665,0.171245) -- (0.96804,0.146342,0.20367) -- (0.96462,0.105599,0.241573) -- (0.956228,0.0704397,0.284019) -- (0.943071,0.0417296,0.329964) -- (0.925473,0.0201758,0.378276) -- (0.903868,0.00630889,0.427765) -- (0.878787,0.000470448,0.477214) -- (0.850848,0.00280419,0.525404) -- (0.82074,0.0132527,0.571149) ; 
\pgfsetxvec{\pgfpoint{1cm}{0cm}}
\pgfsetyvec{\pgfpoint{0cm}{1cm}}

\draw[fill=white] (0,  1) circle (0.05);
\draw[fill=white] (0, -1) circle (0.05);
\end{tikzpicture}%
\label{fig:Bloch_rotation}} \\
\subfloat[conditions on $\alpha_2$ and $\alpha_3$ coefficients]{\includegraphics[width=0.6\columnwidth]{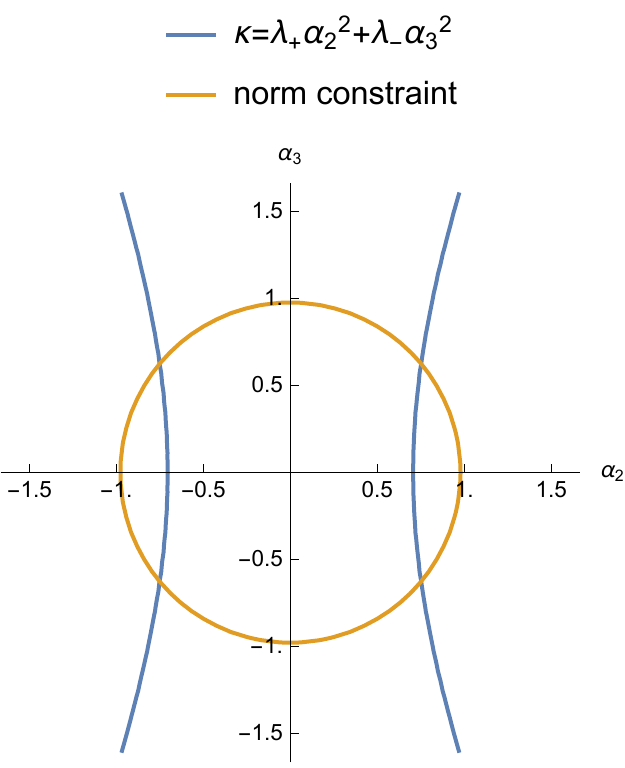}%
\label{fig:roots}}
\caption{(a) The Bloch sphere picture of the time dynamics effected by a Hamiltonian is a classical rotation. For a single observable, the reconstruction of $H$ is not unique: e.g., the two trajectories for $H = \vec{h} \cdot \vec{\sigma}$ and $H' = \vec{h}' \cdot \vec{\sigma}$ result in the same $\braket{Z}$ expectation values. (b) The system of equations to be solved for the reconstruction admits four solutions.}
\end{figure}
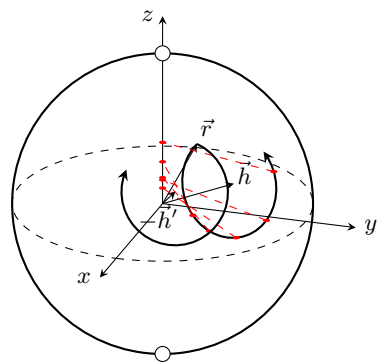
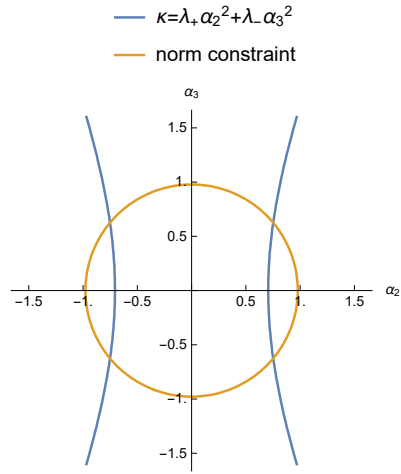

We now describe how to reconstruct a single-qubit Hamiltonian from a time series of measurements. The Bloch sphere picture \cite{nielsen-2010} provides a geometric perspective and insight into single-qubit quantum states and operations. The Bloch vector $\vec{r} \in \R^3$ associated with $\psi \in \C^2$ is defined via the relation
\begin{equation}
\label{eq:bloch_vector_psi}
\ket{\psi}\bra{\psi} = \frac{1}{2}(I + \vec{r} \cdot \vec{\sigma}).
\end{equation}
For pure states as considered here, $\vec{r}$ is a unit vector. We can parametrize any single-qubit Hamiltonian (up to multiples of the identity map, which are irrelevant here) as
\begin{equation}
    \label{eq:single_qubit_H}
    H = \vec{h} \cdot \vec{\sigma}
\end{equation}
with $\vec{h} \in \R^3$. The corresponding time evolution operator is the rotation
\begin{equation}
    U(t) = \e^{-i H t} = \cos(\omega t/2) I - i \sin(\omega t/2) (\vec{v} \cdot \vec{\sigma}) \in \mathrm{SU}(2),
\end{equation}
when representing $\vec{h} = \omega \vec{v} / 2$ by a unit vector $\vec{v} \in \R^3$ and $\omega \in \R$. On the Bloch sphere, this operator corresponds to a classical rotation about $\vec{v}$, as illustrated in Fig.~\ref{fig:Bloch_rotation} and described by Rodrigues' rotation formula ($\theta = \omega t$):
\begin{equation}
\label{eq:rodrigues_rotation}
    \mathsf{U}_{\theta, \vec{v}}\,\vec{r} = \cos(\theta) \vec{r} + \sin(\theta) (\vec{v} \times \vec{r}) + (1 - \cos(\theta)) (\vec{v} \cdot \vec{r}) \,\vec{v}.
\end{equation}
$\mathsf{U}_{\theta, \vec{v}} \in \mathrm{SO}(3)$ is a rotation matrix (parametrized by $\theta$ and $\vec{v}$) applied to $\vec{r}$. In the above context of Takens embedding with $U = U(\Delta t)$, we may equivalently work with $\mathsf{U} = \mathsf{U}_{\omega \Delta t, \vec{v}}$ and matrix powers thereof.

The time trajectory effected by the rotation applied to $\vec{r}$ results in a circle embedded within the Bloch sphere. Knowing the circle would allow to determine $\vec{v}$, and the dynamics on the circle to determine $\omega$. By construction we also know one point on the circle already, namely $\vec{r}$ (the initial condition), but we only have access to the expectation value of an observable $M$ for $t > 0$. In the following, we denote the ``measurement direction'' by $\vec{m} \in \R^3$, i.e., $M$ is parametrized as $M = \vec{m} \cdot \vec{\sigma}$. Algorithm~\ref{alg:reconstruction_alg} facilitates a recovery of $\omega$ and $\vec{v}$ using the projections of the time trajectory on $\vec{m}$. The main idea is to first reconstruct $\omega$ based on the time dependence, and then the unit vector $\vec{v}$.

As caveat, the solution to this problem is not unique, due to the four possible signs of the coefficients $\alpha_2$ and $\alpha_3$ in the algorithm. Fig.~\ref{fig:Bloch_rotation} visualizes how two rotations can generate the same projection onto the $z$-axis, with $\vec{h}'$ resulting from $(\alpha_2, \alpha_3) \to -(\alpha_2, \alpha_3)$. The two equations for $\alpha_2$ and $\alpha_3$ are shown in Fig.~\ref{fig:roots}. (In the example, $\alpha_1 = -0.2051$, and hence the radius of the norm constraint circle is close to $1$.) In order to decide between these distinct solutions, one could perform one further measurement in a different basis, or may use a-priori knowledge about the Hamiltonian.

\begin{algorithm}[H]
\caption{Reconstruction of a single-qubit Hamiltonian (measurement direction $\vec{m} \in \R^3$ with $\vec{m} \nparallel \vec{r}$)}
\label{alg:reconstruction_alg}
\begin{algorithmic}[1]
\Statex \textbf{Input:} Measurement averages $y_q = \vec{m} \cdot (\mathsf{U}_{\omega t_q, \vec{v}} \, \vec{r})$ with $t_q = \Delta t \, \gamma^{q}$ for $q = 0, \dots, 2d$
\Statex \textbf{Output:} Hamiltonian parameters $\omega$ and $\vec{v}$.
\State Find $\omega$ based on the dependency on $\omega t$ in Eq.~\eqref{eq:rodrigues_rotation}:
\[
\omega = \argmin_{\omega > 0} \min_{a, b, c} \sum_{q = 0}^{2d} \, \abs{y_q - a \cos(\omega t_q - b) - c}^2
\]
\State Set $\tilde{y}_q = y_q - \cos(\omega t_q) (\vec{m} \cdot \vec{r})$ for $q = 0, \dots, 2d$ (subtract term which is independent of $\vec{v}$)
\State Find $\alpha_1 = \vec{m} \cdot (\vec{v}\times \vec{r})$ and $\kappa = (\vec{v} \cdot \vec{r}) (\vec{m} \cdot \vec{v})$ via a least squares fit:
\[
\alpha_1, \kappa = \argmin_{\alpha_1, \kappa} \sum_{q=0}^{2d} \abs{\tilde{y}_q - \sin(\omega t_q) \alpha_1 - (1-\cos(\omega t_q)) \kappa}^2
\]
\State Represent $\vec{v}$ with respect to the orthonormal basis
\[
\left\{ \vec{u}_1 = \frac{\vec{r} \times \vec{m}}{\norm{\vec{r} \times \vec{m}}}, \vec{u}_2 =  \frac{\vec{r} + \vec{m}}{\norm{\vec{r} + \vec{m}}}, \vec{u}_3 = \frac{\vec{r} - \vec{m}}{\norm{\vec{r} - \vec{m}}} \right\}:
\]
\[
\vec{v} = \alpha_1 \vec{u}_1 + \alpha_2 \vec{u}_2 + \alpha_3 \vec{u}_3,
\]
with to-be determined coefficients $\alpha_2, \alpha_3 \in \R$. Using that $\vec{u}_1$, $\vec{u}_2$ and $\vec{u}_3$ are eigenvectors of $K = \frac{1}{2} (\vec{m} \otimes \vec{r} + \vec{r} \otimes \vec{m})$ with respective eigenvalues $0$, $\lambda_{+} = \frac{1}{2} (1 + \vec{m} \cdot \vec{r})$ and $\lambda_{-} = -\frac{1}{2} (1 - \vec{m} \cdot \vec{r})$, it follows that
\[
\kappa = \vec{v} \cdot K \vec{v} = \lambda_{+} \alpha_2^2 + \lambda_{-} \alpha_3^2.
\]
Together with the normalization condition
\[
\alpha_1^2 + \alpha_2^2 + \alpha_3^2 = 1,
\]
this leads to $\alpha_2^2 = \kappa - (1 - \alpha_1^2) \lambda_{-}$ and $\alpha_3^2 = 1 - \alpha_1^2 - \alpha_2^2$. Decide between the four possible signs of $\alpha_2, \alpha_3$ using a-priori information of $H$ or one additional measurement in a different basis.
\end{algorithmic}
\end{algorithm}

We remark that the nonlinear optimization in the first step of the algorithm might get trapped in a local minimum, which could be resolved by restarting the optimization. On the other hand, when neglecting uncertainties associated with the measurements, a correct solution is indicated by a zero residual both in steps 1 and 3. We assume that a range of realistic frequencies $\omega$ is known beforehand, such that the Nyquist condition (given the non-uniform sampling points $t_q$) holds \cite{marvasti2001}.

\subsection{Relaxation method}
\label{sec:relaxation}

As straightforward approach for determining a unitary time step matrix $U$ which matches the measurement averages, one could start from the natural parametrization $U = \e^{-i H}$ (with the time step already absorbed into $H$), and then optimize the Hermitian matrix $H$ directly. However, this method encounters the difficulty of a complicated optimization landscape with many local minima, in particular when using gradient descent-based approaches.

Here we discuss an alternative numerical approach tailored towards generic (dense) few-qubit Hamiltonians. As discussed in Sect.~\ref{sec:bloch_single_qubit}, for single qubits a transformation $\psi' = U \psi$ by a unitary matrix $U \in \mathrm{SU}(2)$ can equivalently be described by a spatial rotation in three dimensions on the Bloch sphere: $\vec{r}' = \mathsf{U} r$, with $\mathsf{U} \in \mathrm{SO}(3)$ given by Rodrigues' formula \eqref{eq:rodrigues_rotation}. An equivalent mapping from $U$ to $\mathsf{U}$ is via
\begin{equation}
\label{eq:bloch_unitary_single_qubit}
    \mathsf{U}_{\alpha,\beta} = \frac{1}{2} \tr\!\big[\sigma^{\alpha} U \sigma^{\beta} U^{\dagger}\big]
\end{equation}
for $\alpha, \beta = 1, 2, 3$, where we have used the relation \eqref{eq:bloch_vector_psi} together with the orthogonality relation of the Pauli matrices: $\frac{1}{2} \tr[\sigma^{\alpha} \sigma^{\beta}] = \delta_{\alpha \beta}$. For our purposes, the Bloch picture has the advantage that measurement averages depend linearly on the Bloch vector, and hence on $\mathsf{U}$, e.g., $\braket{\psi' \vert \sigma^z \vert \psi'} = \vec{e}_z \cdot \vec{r}' = \vec{e}_z \cdot (\mathsf{U} r)$. In practice, we first optimize the entries of $\mathsf{U}$ to reproduce the measurement data (under the constraint that $\mathsf{U}$ is an orthogonal matrix), and afterwards find $U$ related to $\mathsf{U}$ via Eq.~\eqref{eq:bloch_unitary_single_qubit}.

Generalization to a larger number of qubits is feasible via tensor products of Pauli and identity matrices (in other words, Pauli strings). The analogue of \eqref{eq:bloch_unitary_single_qubit} for $n$ qubits, with $U \in \mathrm{SU}(2^n)$, is $\mathsf{U} \in \mathrm{SO}(4^n-1)$ with entries 
\begin{equation}
\label{eq:bloch_unitary_n_qubits}
    \mathsf{U}_{\alpha,\beta} = \frac{1}{2^n} \tr\!\big[(\sigma^{\alpha_1} \otimes \cdots \otimes \sigma^{\alpha_n}) U (\sigma^{\beta_1} \otimes \cdots \otimes \sigma^{\beta_n}) U^{\dagger}\big],
\end{equation}
where $\alpha$ and $\beta$ are now index tuples from the set $\{0, 1, 2, 3\}^{\otimes n} \backslash \{ (0, \dots, 0) \}$, using the convention that $\sigma^0$ is the $2 \times 2$ identity matrix. We exclude the tuple $(0, \dots, 0)$ here since it conveys no additional information: the identity matrix is always mapped to itself by unitary conjugation.

Specifying an element of $\mathrm{SO}(4^n - 1)$ involves more free parameters than for a unitary matrix from $\mathrm{SU}(2^n)$ in case $n \ge 2$, thus the optimization might find an orthogonal $\mathsf{U}$ which reproduces the measurement data, but does not originate from a $U \in \mathrm{SU}(2^n)$ via \eqref{eq:bloch_unitary_n_qubits}. We circumvent this representability issue as follows, focusing on the case of two qubits here. Every $U \in \mathrm{SU}(4)$ can be decomposed as \cite{kraus-2001, zhang-2003}
\begin{equation}\label{eq:SU4_decomp}
    U = \big(u^{\mathrm{a}} \otimes u^{\mathrm{b}}\big) R \big(u^{\mathrm{c}} \otimes u^{\mathrm{d}}\big)
\end{equation}
with the ``entanglement'' gate
\begin{equation}
    \label{eq:entanglement_gate}
    R = \e^{-\frac{i}{2} (\theta_1\,\sigma^1 \otimes \sigma^1 + \theta_2\,\sigma^2 \otimes \sigma^2 + \theta_3\,\sigma^3 \otimes \sigma^3)}, \quad \theta_1, \theta_2, \theta_3 \in \R
\end{equation}
and single-qubit unitaries $u^{\mathrm{a}}, u^{\mathrm{b}}, u^{\mathrm{c}}, u^{\mathrm{d}} \in \mathrm{SU}(2)$. The single qubit gates can be handled as before, i.e., represented by orthogonal rotation matrices $\mathsf{u}^{\mathrm{a}}, \mathsf{u}^{\mathrm{b}}, \mathsf{u}^{\mathrm{c}}, \mathsf{u}^{\mathrm{d}} \in \mathrm{SO}(3)$. We find the Bloch representation of $R$ via \eqref{eq:bloch_unitary_n_qubits} (cf.~\cite{gamel-2016, huang-2021}); the parameters $\theta_1, \theta_2, \theta_3$ appear in the matrix entries solely as $\cos(\theta_j)$ and $\sin(\theta_j)$ for $j = 1, 2, 3$. In summary, we express the decomposition \eqref{eq:SU4_decomp} in terms of to-be found orthogonal matrices in the Bloch picture.

After switching to the described Bloch representation, we ``relax'' the condition that an involved (real) matrix $\mathsf{U}$ is orthogonal, by admitting any real matrix, but adding the term
\begin{equation}
    \mathcal{L}_{\text{orth}} = \big\lVert \mathsf{U} \mathsf{U}^T - I \big\rVert_\mathsf{F}^2
\end{equation}
to the overall cost function, where $\norm{\cdot}_\mathsf{F}$ denotes the Frobenius norm. As advantage, we bypass a parametrization of $\mathsf{U}$ to enforce strict orthogonality and avoid local minima in the optimization.

By choosing $\gamma = 2$ in Eq.~\eqref{eq:phi_gamma}, the time steps $t_q = \gamma^q$ are integers, and we can generate corresponding powers of $U$ by defining $\mathsf{U}_0 = \mathsf{U}$, $\mathsf{U}_q = \mathsf{U}_{q-1}^2$ for $q = 1, \dots, 2d$, such that
\begin{equation}
    \e^{-i \gamma^q H} = \mathsf{U}^{\gamma^q} = \mathsf{U}_q \quad \text{for } q = 0, \dots, 2d.
\end{equation}
In our case, the $\mathsf{U}_q$'s are separate matrices, which are set in relation via an additional cost function term
\begin{equation}
    \mathcal{L}_{\text{steps}} = \sum_{q=1}^{2d} \big\lVert \mathsf{U}_q - \mathsf{U}_{q-1}^2 \big\rVert_\mathsf{F}^2.
\end{equation}

For the case of two-qubit Hamiltonians, we additionally substitute two-dimensional vectors $(c_j, s_j)$ for $(\cos(\theta_j), \sin(\theta_j))$, again to avoid local minima. The condition $c_j^2 + s_j^2 = 1$ translates to another penalty term in the overall cost function:
\begin{equation}
    \label{eq:L_theta}
    \mathcal{L}_{\theta} = \sum_{j=1}^3 \abs{c_j^2 + s_j^2 - 1}^2.
\end{equation}

\subsection{Partial (subsystem) measurements}

\begin{figure}[!ht]
\centering
\begin{tikzpicture}[scale=1.500000,x=1pt,y=1pt]
\filldraw[color=white] (0.000000, -7.500000) rectangle (85.000000, 22.500000);
\draw[color=black] (0.000000,15.000000) -- (73.000000,15.000000);
\draw[color=black] (73.000000,14.500000) -- (85.000000,14.500000);
\draw[color=black] (73.000000,15.500000) -- (85.000000,15.500000);
\draw[color=black] (0.000000,0.000000) -- (85.000000,0.000000);
\filldraw[color=white,fill=white] (0.000000,-3.750000) rectangle (-4.000000,18.750000);
\draw[decorate,decoration={brace,amplitude = 4.000000pt},very thick] (0.000000,-3.750000) -- (0.000000,18.750000);
\draw[color=black] (-4.000000,7.500000) node[left] {$\ket{\psi}$};
\draw (18.500000,15.000000) -- (18.500000,0.000000);
\begin{scope}
\draw[fill=white] (18.500000, 7.500000) +(-45.000000:17.677670pt and 19.091883pt) -- +(45.000000:17.677670pt and 19.091883pt) -- +(135.000000:17.677670pt and 19.091883pt) -- +(225.000000:17.677670pt and 19.091883pt) -- cycle;
\clip (18.500000, 7.500000) +(-45.000000:17.677670pt and 19.091883pt) -- +(45.000000:17.677670pt and 19.091883pt) -- +(135.000000:17.677670pt and 19.091883pt) -- +(225.000000:17.677670pt and 19.091883pt) -- cycle;
\draw (18.500000, 7.500000) node {$\e^{-i H t}$};
\end{scope}
\begin{scope}
\draw[fill=white] (49.000000, 15.000000) +(-45.000000:8.485281pt and 8.485281pt) -- +(45.000000:8.485281pt and 8.485281pt) -- +(135.000000:8.485281pt and 8.485281pt) -- +(225.000000:8.485281pt and 8.485281pt) -- cycle;
\clip (49.000000, 15.000000) +(-45.000000:8.485281pt and 8.485281pt) -- +(45.000000:8.485281pt and 8.485281pt) -- +(135.000000:8.485281pt and 8.485281pt) -- +(225.000000:8.485281pt and 8.485281pt) -- cycle;
\draw (49.000000, 15.000000) node {$C$};
\end{scope}
\draw[fill=white] (67.000000, 9.000000) rectangle (79.000000, 21.000000);
\draw[very thin] (73.000000, 15.600000) arc (90:150:6.000000pt);
\draw[very thin] (73.000000, 15.600000) arc (90:30:6.000000pt);
\draw[->,>=stealth] (73.000000, 9.600000) -- +(80:10.392305pt);
\end{tikzpicture}
\caption{Minimal example for the time-delayed partial (subsystem) measurement scenario. The goal is to reconstruct the (unknown) Hamiltonian $H$. We assume that one can prepare two different initial states, and can select a measurement basis via the gate $C$.}
\label{fig:partial_meas_circuit}
\end{figure}
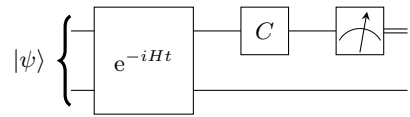

An intriguing possibility of the time-delay measurements is the identification of the overall Hamiltonian based on measurements restricted to a subsystem. This scenario will actually require the preparation of more than one exactly known initial state. As minimal example, consider a two-qubit system, the choice between two initial states, being able to select a measurement basis via the gate $C$, and time-delayed measurements on one of the qubits while the other qubit is inaccessible, as illustrated in Fig.~\ref{fig:partial_meas_circuit}.

For the numerical experiment shown in Fig.~\ref{fig:two_qubits_partial_meas} below, we use $X$-, $Y$- and $Z$-basis measurements on the top qubit and minimize the mean squared error between the observed measurement averages and the prediction based on a general Ansatz for the (traceless) Hamiltonian, i.e., $15$ real parameters.

\subsection{Lattice system with local interactions}

Finally, we consider quantum systems defined on a lattice, and assume an Ising-type Hamiltonian as in Eq.~\eqref{eq:HIsing} (instead of a generic Hamiltonian) to avoid an exponential growth of the number of parameters.

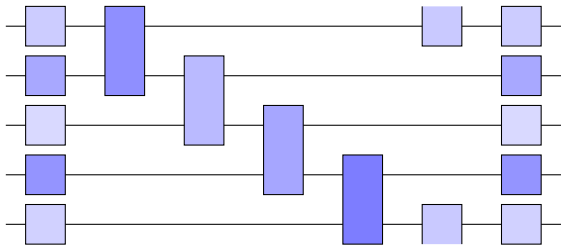
\begin{figure}[!ht]
\centering
\begin{tikzpicture}[scale=1.250000,x=1pt,y=1pt]
\filldraw[color=white] (0.000000, -7.500000) rectangle (168.000000, 67.500000);
\draw[color=black] (0.000000,60.000000) -- (168.000000,60.000000);
\draw[color=black] (0.000000,45.000000) -- (168.000000,45.000000);
\draw[color=black] (0.000000,30.000000) -- (168.000000,30.000000);
\draw[color=black] (0.000000,15.000000) -- (168.000000,15.000000);
\draw[color=black] (0.000000,0.000000) -- (168.000000,0.000000);
\begin{scope}
\draw[fill=blue!20!white] (12.000000, 60.000000) +(-45.000000:8.485281pt and 8.485281pt) -- +(45.000000:8.485281pt and 8.485281pt) -- +(135.000000:8.485281pt and 8.485281pt) -- +(225.000000:8.485281pt and 8.485281pt) -- cycle;
\clip (12.000000, 60.000000) +(-45.000000:8.485281pt and 8.485281pt) -- +(45.000000:8.485281pt and 8.485281pt) -- +(135.000000:8.485281pt and 8.485281pt) -- +(225.000000:8.485281pt and 8.485281pt) -- cycle;
\draw (12.000000, 60.000000) node {{}};
\end{scope}
\begin{scope}
\draw[fill=blue!34!white] (12.000000, 45.000000) +(-45.000000:8.485281pt and 8.485281pt) -- +(45.000000:8.485281pt and 8.485281pt) -- +(135.000000:8.485281pt and 8.485281pt) -- +(225.000000:8.485281pt and 8.485281pt) -- cycle;
\clip (12.000000, 45.000000) +(-45.000000:8.485281pt and 8.485281pt) -- +(45.000000:8.485281pt and 8.485281pt) -- +(135.000000:8.485281pt and 8.485281pt) -- +(225.000000:8.485281pt and 8.485281pt) -- cycle;
\draw (12.000000, 45.000000) node {{}};
\end{scope}
\begin{scope}
\draw[fill=blue!15!white] (12.000000, 30.000000) +(-45.000000:8.485281pt and 8.485281pt) -- +(45.000000:8.485281pt and 8.485281pt) -- +(135.000000:8.485281pt and 8.485281pt) -- +(225.000000:8.485281pt and 8.485281pt) -- cycle;
\clip (12.000000, 30.000000) +(-45.000000:8.485281pt and 8.485281pt) -- +(45.000000:8.485281pt and 8.485281pt) -- +(135.000000:8.485281pt and 8.485281pt) -- +(225.000000:8.485281pt and 8.485281pt) -- cycle;
\draw (12.000000, 30.000000) node {{}};
\end{scope}
\begin{scope}
\draw[fill=blue!42!white] (12.000000, 15.000000) +(-45.000000:8.485281pt and 8.485281pt) -- +(45.000000:8.485281pt and 8.485281pt) -- +(135.000000:8.485281pt and 8.485281pt) -- +(225.000000:8.485281pt and 8.485281pt) -- cycle;
\clip (12.000000, 15.000000) +(-45.000000:8.485281pt and 8.485281pt) -- +(45.000000:8.485281pt and 8.485281pt) -- +(135.000000:8.485281pt and 8.485281pt) -- +(225.000000:8.485281pt and 8.485281pt) -- cycle;
\draw (12.000000, 15.000000) node {{}};
\end{scope}
\begin{scope}
\draw[fill=blue!18!white] (12.000000, 0.000000) +(-45.000000:8.485281pt and 8.485281pt) -- +(45.000000:8.485281pt and 8.485281pt) -- +(135.000000:8.485281pt and 8.485281pt) -- +(225.000000:8.485281pt and 8.485281pt) -- cycle;
\clip (12.000000, 0.000000) +(-45.000000:8.485281pt and 8.485281pt) -- +(45.000000:8.485281pt and 8.485281pt) -- +(135.000000:8.485281pt and 8.485281pt) -- +(225.000000:8.485281pt and 8.485281pt) -- cycle;
\draw (12.000000, 0.000000) node {{}};
\end{scope}
\draw (36.000000,60.000000) -- (36.000000,45.000000);
\begin{scope}
\draw[fill=blue!44!white] (36.000000, 52.500000) +(-45.000000:8.485281pt and 19.091883pt) -- +(45.000000:8.485281pt and 19.091883pt) -- +(135.000000:8.485281pt and 19.091883pt) -- +(225.000000:8.485281pt and 19.091883pt) -- cycle;
\clip (36.000000, 52.500000) +(-45.000000:8.485281pt and 19.091883pt) -- +(45.000000:8.485281pt and 19.091883pt) -- +(135.000000:8.485281pt and 19.091883pt) -- +(225.000000:8.485281pt and 19.091883pt) -- cycle;
\draw (36.000000, 52.500000) node {{}};
\end{scope}
\draw (60.000000,45.000000) -- (60.000000,30.000000);
\begin{scope}
\draw[fill=blue!27!white] (60.000000, 37.500000) +(-45.000000:8.485281pt and 19.091883pt) -- +(45.000000:8.485281pt and 19.091883pt) -- +(135.000000:8.485281pt and 19.091883pt) -- +(225.000000:8.485281pt and 19.091883pt) -- cycle;
\clip (60.000000, 37.500000) +(-45.000000:8.485281pt and 19.091883pt) -- +(45.000000:8.485281pt and 19.091883pt) -- +(135.000000:8.485281pt and 19.091883pt) -- +(225.000000:8.485281pt and 19.091883pt) -- cycle;
\draw (60.000000, 37.500000) node {{}};
\end{scope}
\draw (84.000000,30.000000) -- (84.000000,15.000000);
\begin{scope}
\draw[fill=blue!35!white] (84.000000, 22.500000) +(-45.000000:8.485281pt and 19.091883pt) -- +(45.000000:8.485281pt and 19.091883pt) -- +(135.000000:8.485281pt and 19.091883pt) -- +(225.000000:8.485281pt and 19.091883pt) -- cycle;
\clip (84.000000, 22.500000) +(-45.000000:8.485281pt and 19.091883pt) -- +(45.000000:8.485281pt and 19.091883pt) -- +(135.000000:8.485281pt and 19.091883pt) -- +(225.000000:8.485281pt and 19.091883pt) -- cycle;
\draw (84.000000, 22.500000) node {{}};
\end{scope}
\draw (108.000000,15.000000) -- (108.000000,0.000000);
\begin{scope}
\draw[fill=blue!51!white] (108.000000, 7.500000) +(-45.000000:8.485281pt and 19.091883pt) -- +(45.000000:8.485281pt and 19.091883pt) -- +(135.000000:8.485281pt and 19.091883pt) -- +(225.000000:8.485281pt and 19.091883pt) -- cycle;
\clip (108.000000, 7.500000) +(-45.000000:8.485281pt and 19.091883pt) -- +(45.000000:8.485281pt and 19.091883pt) -- +(135.000000:8.485281pt and 19.091883pt) -- +(225.000000:8.485281pt and 19.091883pt) -- cycle;
\draw (108.000000, 7.500000) node {{}};
\end{scope}
\begin{scope}
\draw[fill=blue!21!white] (132.000000, 0.000000) +(-45.000000:8.485281pt and 8.485281pt) -- +(45.000000:8.485281pt and 8.485281pt) -- +(135.000000:8.485281pt and 8.485281pt) -- +(225.000000:8.485281pt and 8.485281pt);
\clip (132.000000, 0.000000) +(-45.000000:8.485281pt and 8.485281pt) -- +(45.000000:8.485281pt and 8.485281pt) -- +(135.000000:8.485281pt and 8.485281pt) -- +(225.000000:8.485281pt and 8.485281pt) -- cycle;
\draw (132.000000, 0.000000) node {{}};
\end{scope}
\begin{scope}
\draw[fill=blue!21!white] (132.000000, 60.000000) +(135.000000:8.485281pt and 8.485281pt) -- +(225.000000:8.485281pt and 8.485281pt) -- +(-45.000000:8.485281pt and 8.485281pt) -- +(45.000000:8.485281pt and 8.485281pt);
\clip (132.000000, 60.000000) +(-45.000000:8.485281pt and 8.485281pt) -- +(45.000000:8.485281pt and 8.485281pt) -- +(135.000000:8.485281pt and 8.485281pt) -- +(225.000000:8.485281pt and 8.485281pt) -- cycle;
\draw (132.000000, 60.000000) node {{}};
\end{scope}
\begin{scope}
\draw[fill=blue!20!white] (156.000000, 60.000000) +(-45.000000:8.485281pt and 8.485281pt) -- +(45.000000:8.485281pt and 8.485281pt) -- +(135.000000:8.485281pt and 8.485281pt) -- +(225.000000:8.485281pt and 8.485281pt) -- cycle;
\clip (156.000000, 60.000000) +(-45.000000:8.485281pt and 8.485281pt) -- +(45.000000:8.485281pt and 8.485281pt) -- +(135.000000:8.485281pt and 8.485281pt) -- +(225.000000:8.485281pt and 8.485281pt) -- cycle;
\draw (156.000000, 60.000000) node {{}};
\end{scope}
\begin{scope}
\draw[fill=blue!34!white] (156.000000, 45.000000) +(-45.000000:8.485281pt and 8.485281pt) -- +(45.000000:8.485281pt and 8.485281pt) -- +(135.000000:8.485281pt and 8.485281pt) -- +(225.000000:8.485281pt and 8.485281pt) -- cycle;
\clip (156.000000, 45.000000) +(-45.000000:8.485281pt and 8.485281pt) -- +(45.000000:8.485281pt and 8.485281pt) -- +(135.000000:8.485281pt and 8.485281pt) -- +(225.000000:8.485281pt and 8.485281pt) -- cycle;
\draw (156.000000, 45.000000) node {{}};
\end{scope}
\begin{scope}
\draw[fill=blue!15!white] (156.000000, 30.000000) +(-45.000000:8.485281pt and 8.485281pt) -- +(45.000000:8.485281pt and 8.485281pt) -- +(135.000000:8.485281pt and 8.485281pt) -- +(225.000000:8.485281pt and 8.485281pt) -- cycle;
\clip (156.000000, 30.000000) +(-45.000000:8.485281pt and 8.485281pt) -- +(45.000000:8.485281pt and 8.485281pt) -- +(135.000000:8.485281pt and 8.485281pt) -- +(225.000000:8.485281pt and 8.485281pt) -- cycle;
\draw (156.000000, 30.000000) node {{}};
\end{scope}
\begin{scope}
\draw[fill=blue!42!white] (156.000000, 15.000000) +(-45.000000:8.485281pt and 8.485281pt) -- +(45.000000:8.485281pt and 8.485281pt) -- +(135.000000:8.485281pt and 8.485281pt) -- +(225.000000:8.485281pt and 8.485281pt) -- cycle;
\clip (156.000000, 15.000000) +(-45.000000:8.485281pt and 8.485281pt) -- +(45.000000:8.485281pt and 8.485281pt) -- +(135.000000:8.485281pt and 8.485281pt) -- +(225.000000:8.485281pt and 8.485281pt) -- cycle;
\draw (156.000000, 15.000000) node {{}};
\end{scope}
\begin{scope}
\draw[fill=blue!18!white] (156.000000, 0.000000) +(-45.000000:8.485281pt and 8.485281pt) -- +(45.000000:8.485281pt and 8.485281pt) -- +(135.000000:8.485281pt and 8.485281pt) -- +(225.000000:8.485281pt and 8.485281pt) -- cycle;
\clip (156.000000, 0.000000) +(-45.000000:8.485281pt and 8.485281pt) -- +(45.000000:8.485281pt and 8.485281pt) -- +(135.000000:8.485281pt and 8.485281pt) -- +(225.000000:8.485281pt and 8.485281pt) -- cycle;
\draw (156.000000, 0.000000) node {{}};
\end{scope}
\end{tikzpicture}
\caption{One time step $\Delta t$ of the quantum dynamics based on Strang splitting into interaction and local field terms, see Eq.~\eqref{eq:H_int_loc}, illustrated for a one-dimensional lattice with $n = 5$ sites. The single qubit gates on the left and right are rotation gates $\exp(-i \Delta t \, \vec{h}_j \cdot \vec{\sigma}_j/2)$ for the $j$-th qubit, and the two-qubit gates are $\exp(-i \Delta t J_{j,j+1} \sigma^z_j \sigma^z_{j+1})$. The different shades indicate different parameters at each site. The ordering of the two-qubit interaction gates among each other is not relevant since they pairwise commute.}
\label{fig:strang_int_local_circuit}
\end{figure}

For the purpose of reconstructing the Hamiltonian parameters, the Schr\"odinger time evolution can be well approximated via a second order Strang splitting method \cite{mclachlan-2002}, i.e., the Time-Evolving Block Decimation (TEBD) algorithm \cite{vidal2004}. Interpreting the resulting layout of single- and two-site unitary operators as forming a quantum circuit, the reconstruction task amounts to a variational circuit optimization to reproduce the reference measurement averages. Specifically, when considering the interaction and local field parts of the Hamiltonian
\begin{equation}
\label{eq:H_int_loc}
H_{\text{int}} = - \sum_{\langle j, \ell \rangle} J_{j,\ell} \, \sigma^z_j \sigma^z_\ell, \quad H_{\text{loc}} = - \sum_{j \in \Lambda} \vec{h}_j \cdot \vec{\sigma}_j
\end{equation}
by themselves, the individual terms in each of the sums pairwise commute. Fig.~\ref{fig:strang_int_local_circuit} illustrates the corresponding quantum circuit for a one-dimensional lattice; the constructing works for higher-dimensional lattices as well.

\section{Numerical experiments}

Here we present numerical experiments for the methods described in Sect.~\ref{sec:methods}.

\subsection{Exact reconstruction of a single-qubit Hamiltonian}

\begin{figure}[!ht]
\centering
\subfloat[cosine fit to find $\omega$]{\includegraphics[width=0.8\columnwidth]{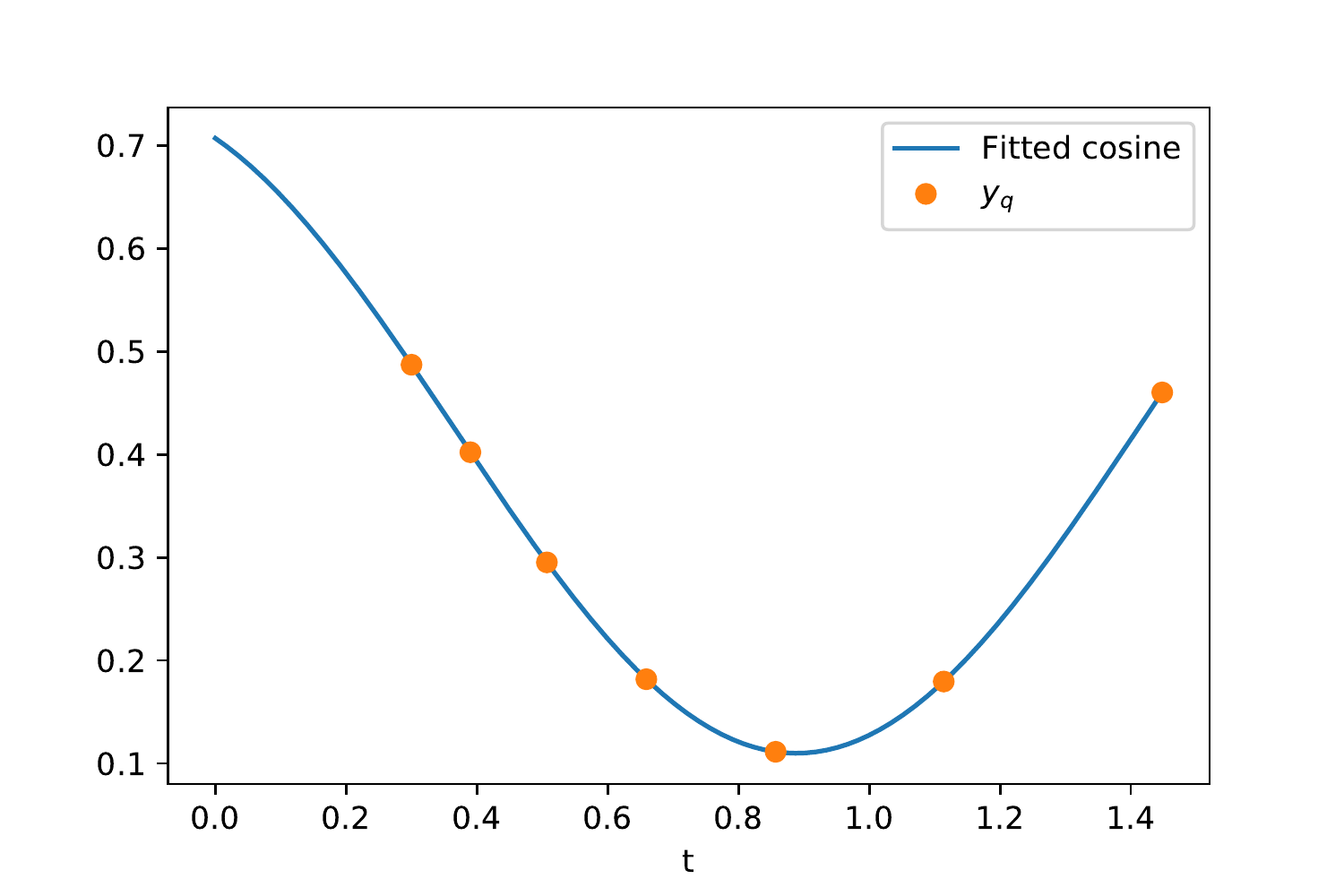}%
\label{fig:cos-fit}} \\
\subfloat[function fit to find $\alpha_1$ and $\kappa$]{\includegraphics[width=0.8\columnwidth]{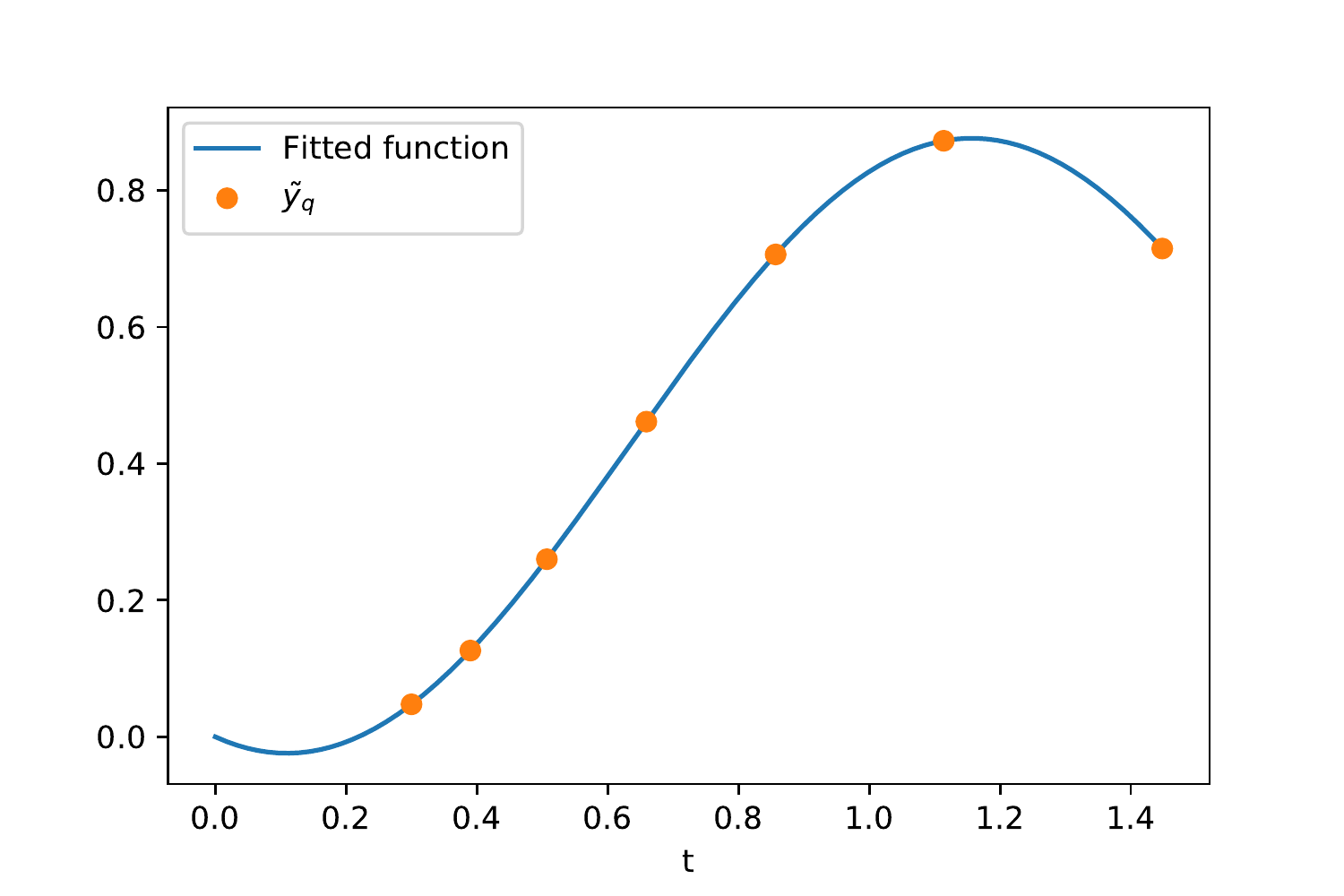}%
\label{fig:sincos-fit}}
\caption{(a) step 1 and (b) step 3 of Algorithm~\ref{alg:reconstruction_alg} using $Z$-basis measurements, for the time evolution shown in Fig.~\ref{fig:Bloch_rotation}.}
\label{fig:single-qubit-exact}
\end{figure}

With the algorithm described in Sec.~\ref{sec:bloch_single_qubit} we are able to reconstruct a single-qubit Hamiltonian up to numerical precision. Fig.~\ref{fig:single-qubit-exact} shows the results obtained for the $\omega$ and least squares optimizations. Following the Takens framework, we use $2d + 1$ time points, where $d = 3$ is the number of Hamiltonian parameters to be reconstructed. Specifically, we have used the time points $t_q = 0.3 \times (1.3)^q$ for $q = 0, \dots, 6$ here. With this, we are able to find $\omega$, $\alpha_1$ and $\kappa$ up to an error of $\sim10^{-15}$. Thus $7$ measurements, plus a further measurement in a different basis to distinguish between the possible solutions due to symmetry, are sufficient for the reconstruction.

\subsection{Relaxation method}

In this subsection we consider a generic single- or two-qubit Hamiltonian $H$. Specifically, we draw the coefficients of $H$ with respect to the Pauli basis from the standard normal (Gaussian) distribution.

\begin{figure}[!htp]
\centering
\subfloat[direct parameter fitting (single qubit)]{\includegraphics[width=0.8\columnwidth]{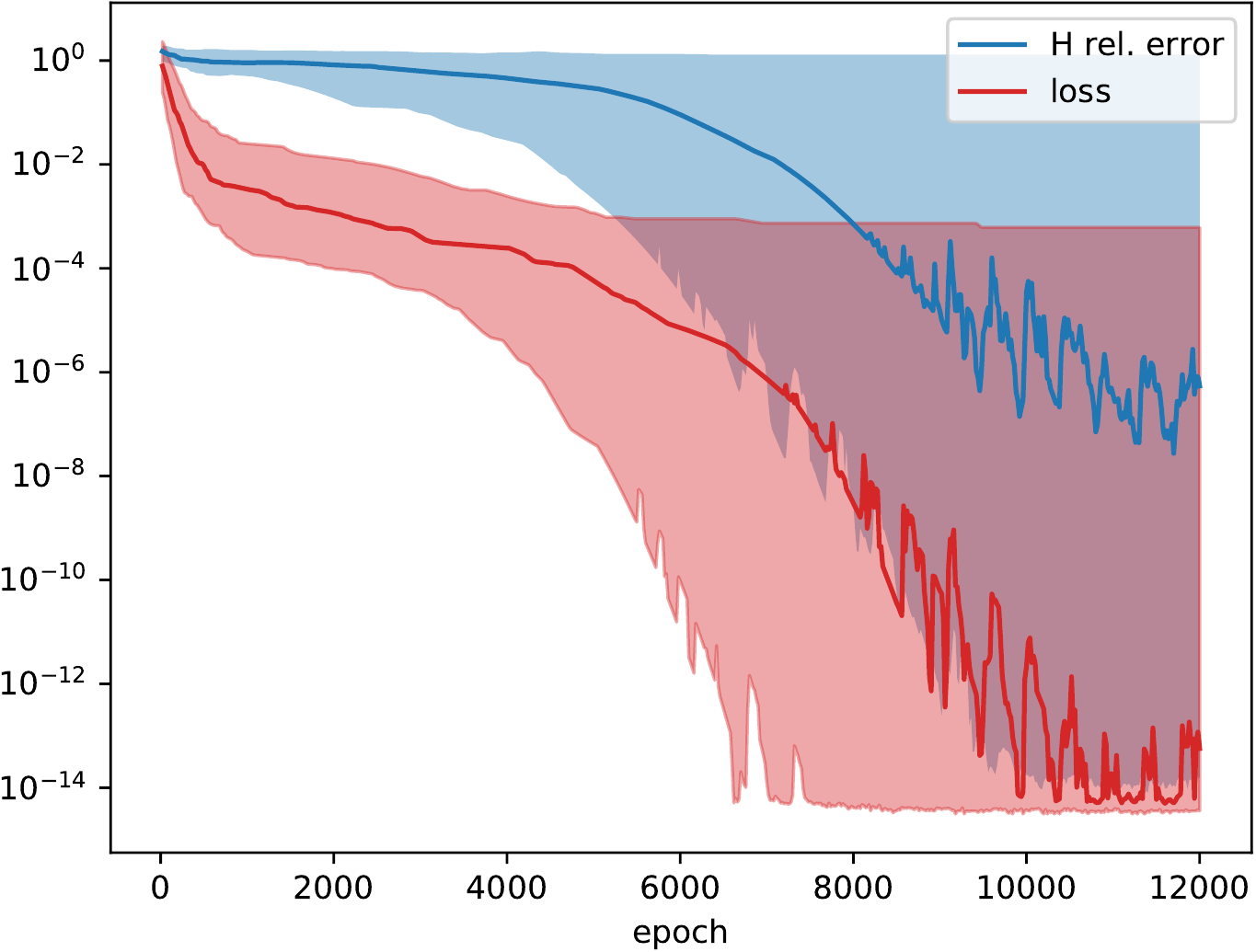}%
\label{fig:reconstruct_dense_single_qubit_H}} \\[1.5\baselineskip]
\subfloat[``relaxation'' method (single qubit)]{\includegraphics[width=0.8\columnwidth]{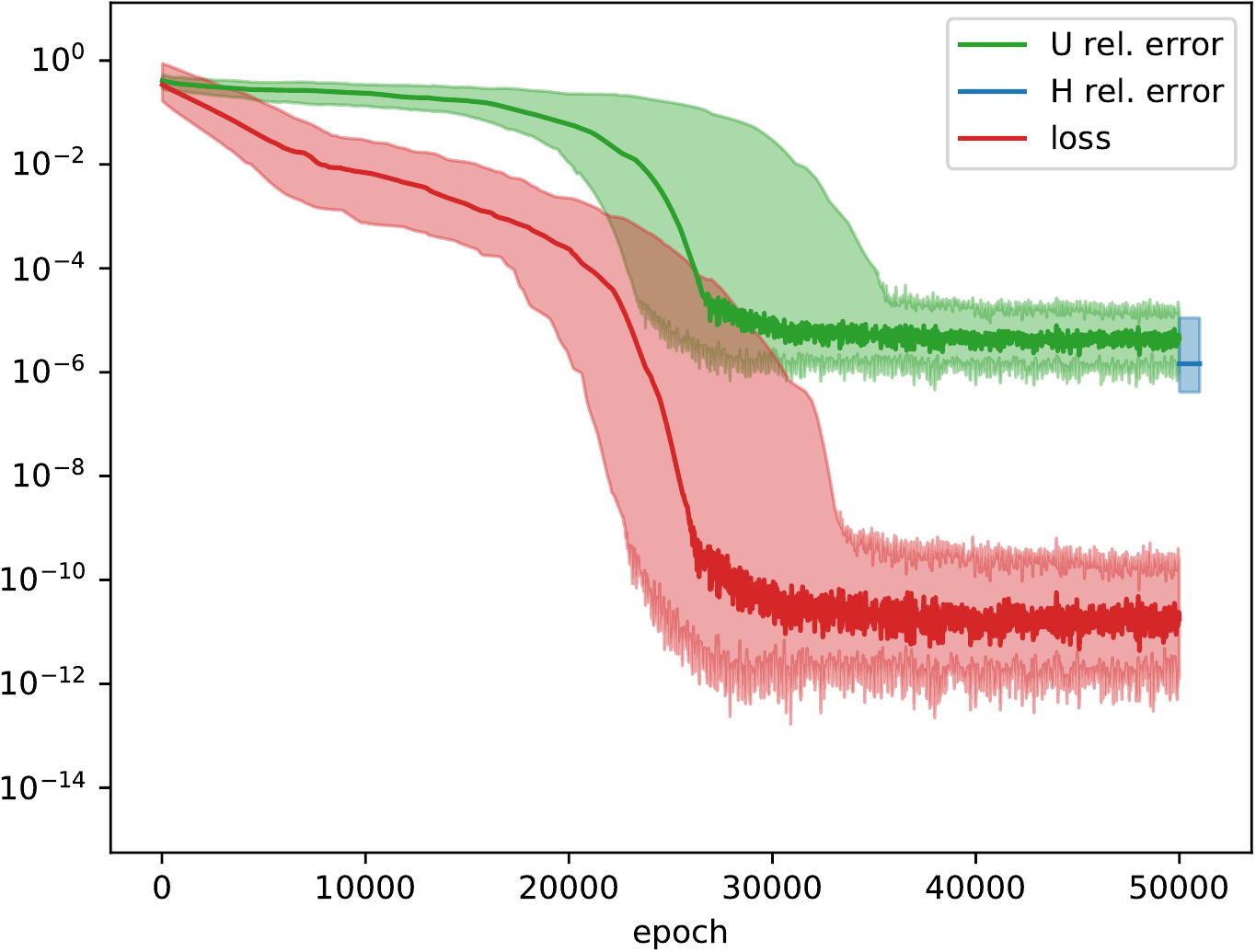}%
\label{fig:reconstruct_dense_single_qubit_U_relax}} \\[1.5\baselineskip]
\subfloat[``relaxation'' method (two qubits)]{\includegraphics[width=0.8\columnwidth]{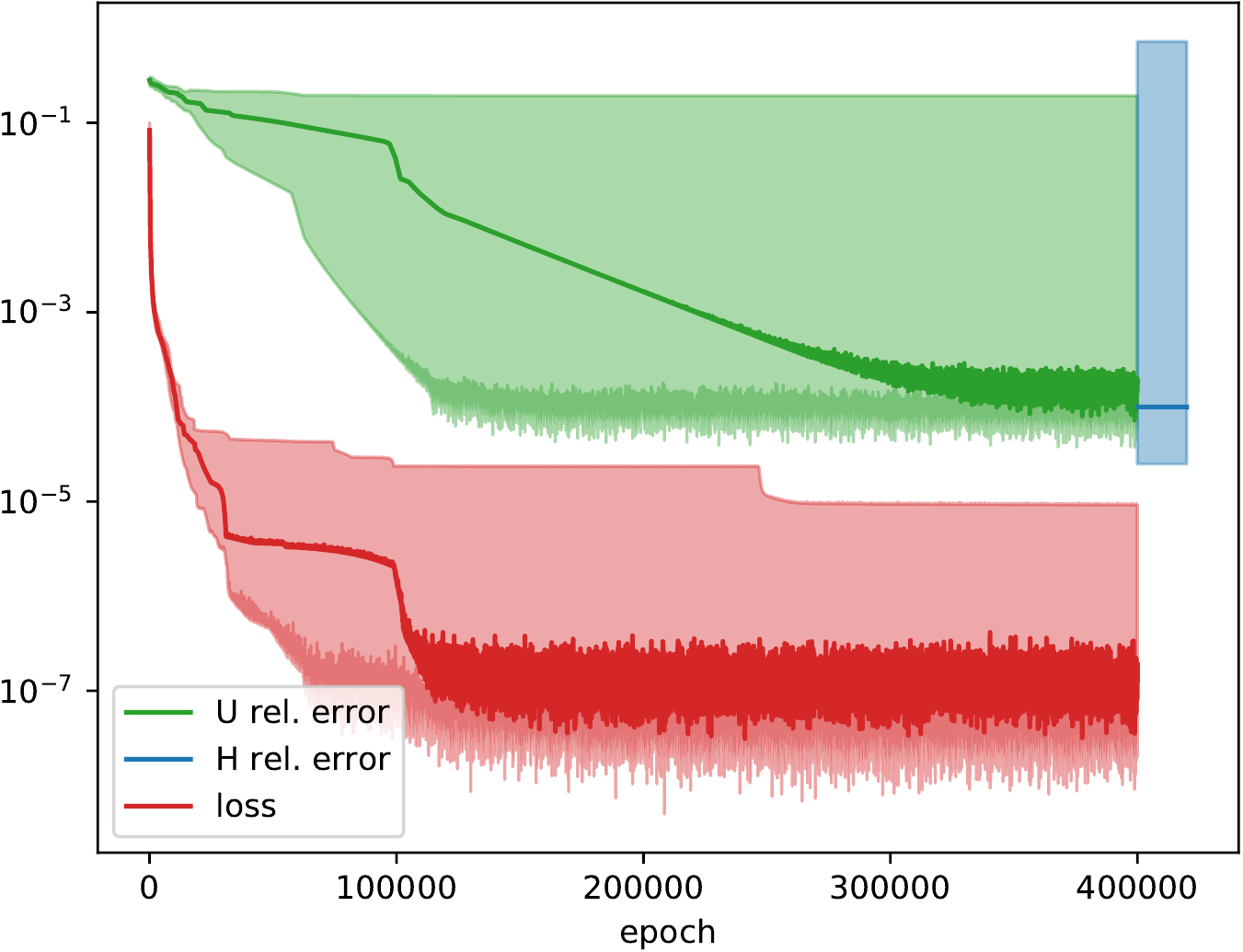}%
\label{fig:reconstruct_dense_two_qubit}}
\caption{(a) and (b): Single-qubit Hamiltonian reconstruction based on four $Z$- and four $X$-basis measurements, (a) by direct fitting of the Hamiltonian parameters to the measurement data, and (b) using the ``relaxation'' procedure (Sect.~\ref{sec:relaxation}), with the actual Hamiltonian parameters determined from $\mathsf{U}$ at the end (blue line at the rightmost section of the plot). Solid lines show the median, and lighter shades are $25\%$ quantiles. (c) ``Relaxation procedure'' for two-qubit Hamiltonian reconstruction, working with the Bloch picture of the representation~\eqref{eq:SU4_decomp} to find the time step matrix $\mathsf{U}$, and then determining a two-qubit Hamiltonian giving rise to this $\mathsf{U}$.}
\end{figure}

We first focus on a single-qubit $2 \times 2$ Hamiltonian. Fig.~\ref{fig:reconstruct_dense_single_qubit_H} shows the result of directly fitting the vector $\vec{h} \in \R^3$ (see Eq.~\eqref{eq:single_qubit_H}), with the KL divergence between the predicted and actual measurement probabilities as cost function. This approach is compared to the ``relaxation'' procedure (described in Sect.~\ref{sec:relaxation}) in Fig.~\ref{fig:reconstruct_dense_single_qubit_U_relax}, using $\gamma = 2$ and $\Delta t = 0.1$. The manifold for the Takens embedding has dimension $d = 3$ for a single qubit, and hence $2d + 1 = 7$ time points should be used. However, we face the difficulty that the Hamiltonian is not uniquely specified solely by Pauli-$Z$ measurements according to the discussion in Sect.~\ref{sec:bloch_single_qubit}. For this reason, we include Pauli-$X$ measurement data as well, and reduce the number of time points to $4$ (to compensate for this additional source of data). For both versions we use the Adam optimizer \cite{kingma-2015}. The direct fitting method might get trapped in a local minimum and hence not be able to find the ground-truth vector $\vec{h}$, which leads to a large variation around the median. This issue is ameliorated by the relaxation method.

In Fig.~\ref{fig:reconstruct_dense_two_qubit} visualizes the loss function and relative errors when applying the relaxation procedure to reconstruct the unitary time step matrix and corresponding Hamiltonian of a two-qubit system ($d = 15$ real parameters). Specifically, for parameter optimization we express Eq.~\eqref{eq:SU4_decomp} using the Bloch picture, i.e., in terms of orthogonal rotation matrices for the single-qubit unitaries, and likewise using the Bloch picture analogue of the entanglement gate \eqref{eq:entanglement_gate}, parametrized by two-dimensional vectors $(c_j, s_j)$ for $(\cos(\theta_j), \sin(\theta_j))$. The condition $c_j^2 + s_j^2 = 1$ translates to another penalty term in the overall cost function, see Eq.~\eqref{eq:L_theta}. We set $\Delta t = 0.05$, $\gamma = 2$ and use $6$ time points for this experiment. The reference measurement data at each time point are the expectation values of the nine observables $\{ I \otimes \sigma^{\alpha}, \sigma^{\alpha} \otimes I, \sigma^{\alpha} \otimes \sigma^{\alpha}\}_{\alpha \in \{ x, y, z \}}$. We use $54$ measurement data points in total (instead of $2d + 1 = 31$) since we found that the additional information improves the reconstruction. As last step (after the main optimization), we find the Hamiltonian entries giving rise to the computed $\mathsf{U}$ via the Broyden-Fletcher-Goldfarb-Shanno (BFGS) algorithm.

According to Fig.~\ref{fig:reconstruct_dense_two_qubit}, the median relative errors of $\mathsf{U}$ and the Hamiltonian are below $10^{-3}$, but there are still cases where the method cannot find the reference solution at all, even tough the loss value is small. An explanation could be that $\mathsf{U}$ and $H$ are not uniquely determined. We leave a clarification of this point for future work.

\subsection{Partial (subsystem) measurements}\label{partial-meas}

We study the scenario depicted in Fig.~\ref{fig:partial_meas_circuit}, namely performing measurements solely on one out of two qubits. The observables are the three Pauli gates here. For the reconstruction to work, we require that two different, precisely characterized initial states can be prepared, which we choose at random for the numerical experiments.

The ground-truth Hamiltonian is similarly constructed at random, by drawing standard normal-distributed coefficients of Pauli strings, which in sum form the Hamiltonian. We use the time step $\Delta t = \frac{1}{5}$ and $\gamma = 1.15$ here, and have heuristically found $12$ time-delayed measurements (for each measurement basis) as viable to reliably reconstruct the Hamiltonian. The loss function is the mean squared error between the model prediction for the measurement averages and the ground-truth values. To avoid getting trapped in local minima during the numerical optimization, we use the best out of 10 attempts (in terms of the loss function) with different random starting points.

\begin{figure}[!ht]
\subfloat[example trajectory]{\includegraphics[width=0.8\columnwidth]{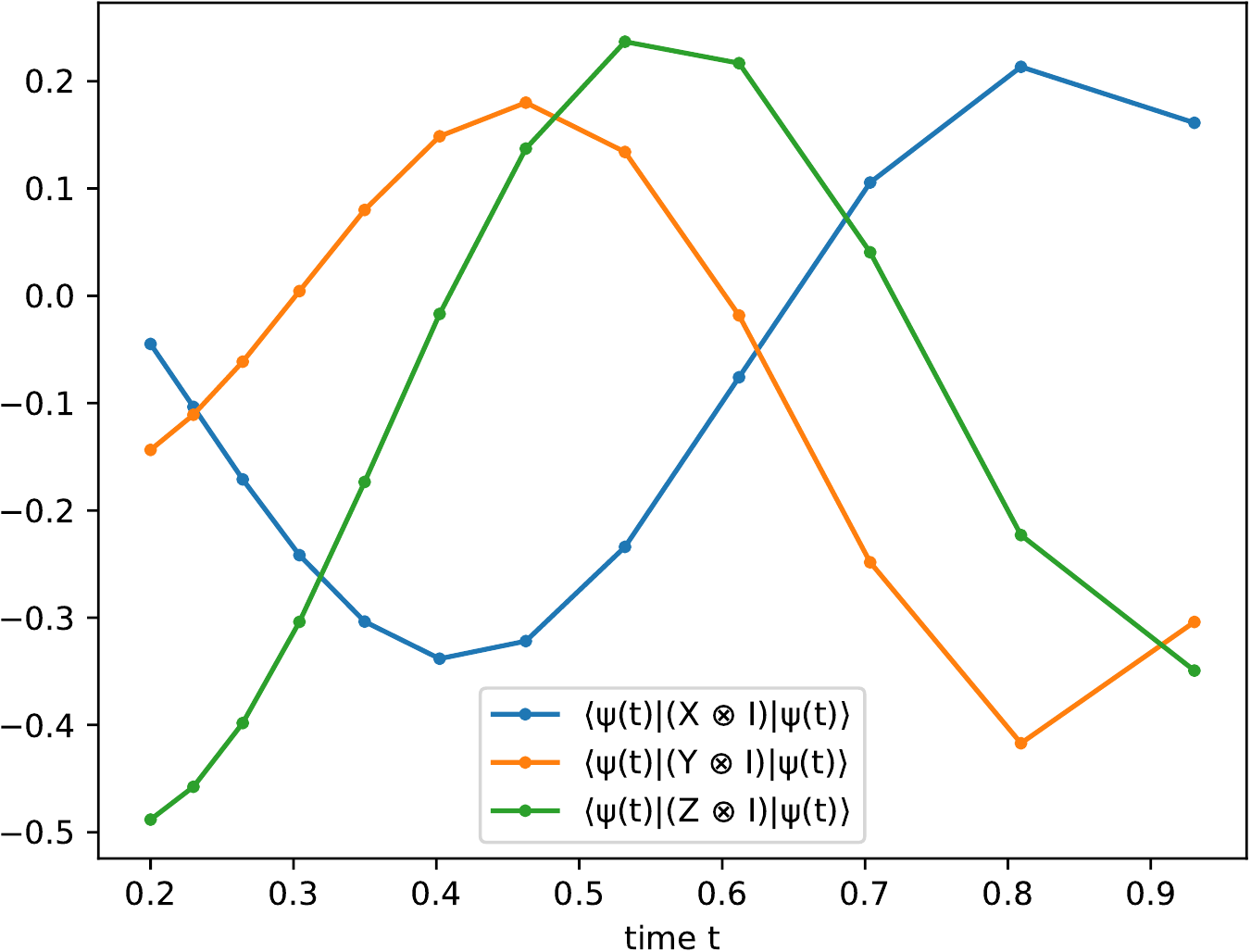}%
\label{fig:two_qubits_partial_meas_trajectory}} \\
\subfloat[reconstruction error and loss function]{\includegraphics[width=0.8\columnwidth]{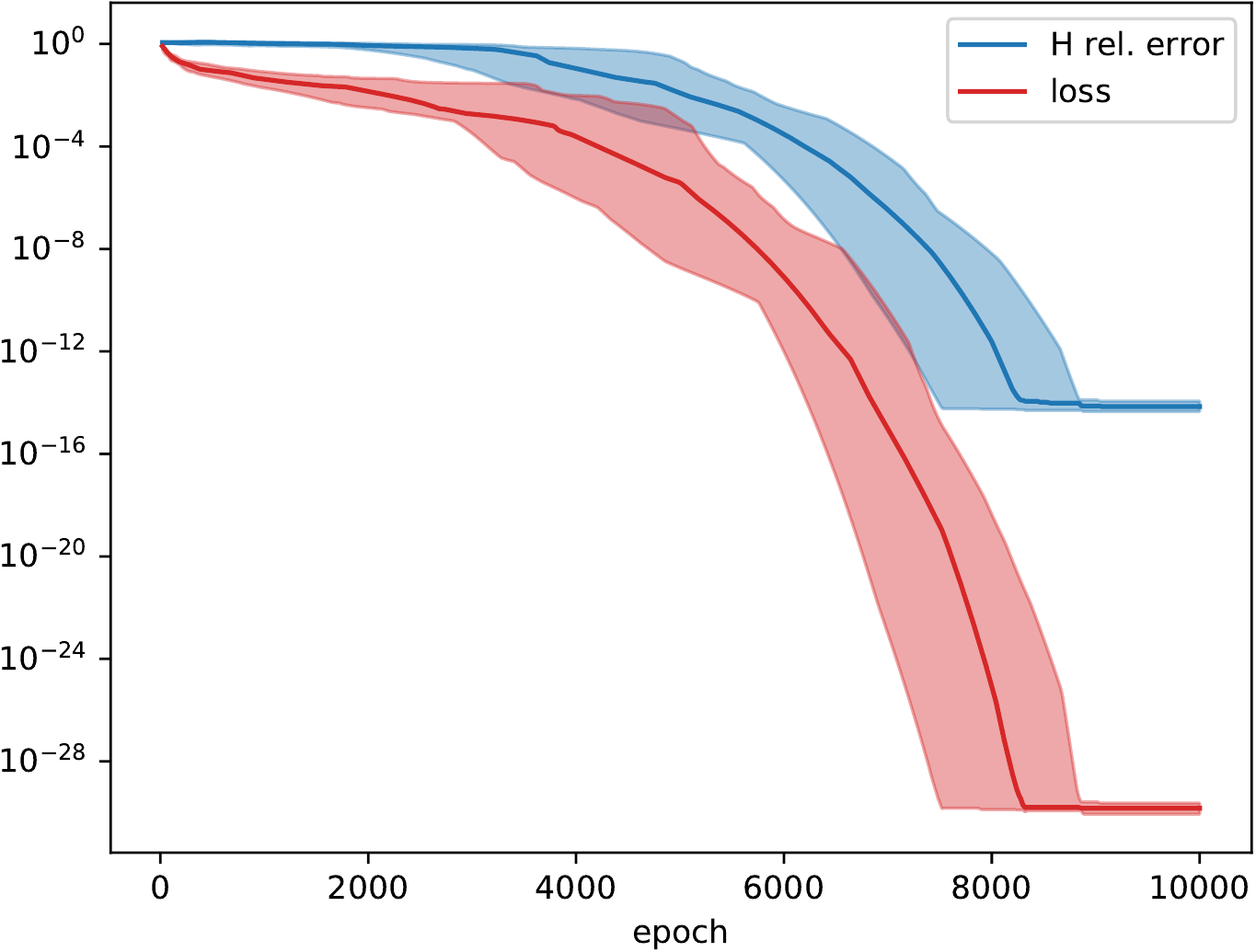}%
\label{fig:two_qubits_partial_meas_training}}
\caption{Numerical reconstruction of a generic two-qubit Hamiltonian based on time-delayed measurements of only one of the qubits, cf.~Fig.~\ref{fig:partial_meas_circuit}. (a) Exemplary measurement time trajectory of the operators $\sigma^x \otimes I$, $\sigma^y \otimes I$ and $\sigma^z \otimes I$. Two random initial states and their respective trajectories are used as input to the reconstruction algorithm. (b) Reconstruction error and loss function based on $20$ random realizations.}
\label{fig:two_qubits_partial_meas}
\end{figure}

The specific observables are the three Pauli gates applied to the first qubit. Fig.~\ref{fig:two_qubits_partial_meas_trajectory} shows exemplary measurement time trajectories, and Fig.~\ref{fig:two_qubits_partial_meas_training} the reconstruction error and loss function (median and $25\%$ quantiles based on $20$ random realizations of the overall setup). One observes that a reliable and precise reconstruction of the Hamiltonian (even up to numerical rounding errors) is possible in principle, when neglecting inaccuracies of the measurement process.

We remark that we have used only a single initial state and less time points for the ``relaxation'' method (see Fig.~\ref{fig:reconstruct_dense_two_qubit}), which can explain the seemingly larger errors there.

\subsection{Hamiltonian on a lattice with local interactions}

We first consider the case of a Hamiltonian \eqref{eq:HIsing} with \emph{uniform} parameters (not depending on the lattice site), i.e.,
\begin{equation}
\label{eq:HIsing_uniform}
H = - \sum_{\langle j, \ell \rangle} J \, \sigma^z_j \sigma^z_\ell - \sum_{j \in \Lambda} \vec{h} \cdot \vec{\sigma}_j
\end{equation}
with $J \in \mathbb{R}$ and $\vec{h} \in \mathbb{R}^3$. Thus the task consists of reconstructing $4$ real numbers. The ground-truth values for the following experiment are $J = 1$ and $\vec{h} = (0.5, -0.8, 1.1)$. $\Lambda$ is a $3 \times 4$ lattice with periodic boundary conditions, and the initial state is a single wavefunction with complex random entries (independently normally distributed). We compute the time-evolved quantum state $\psi(t)$ via the KrylovKit Julia package \cite{krylovkit}, and use the squared entries of $\psi(t)$ as reference Born measurement averages.

For the purpose of reconstructing $J$ and $\vec{h}$, we approximate a time step via a variational quantum circuit as shown in Fig.~\ref{fig:strang_int_local_circuit}, where the single- and two-qubit gates share their to-be optimized parameters $\tilde{J}$ and $\vec{\tilde{h}}$. For simplicity, we use three uniform time points $\Delta t$, $2 \Delta t$, $3 \Delta t$ with $\Delta t = 0.2$ (instead of time points $t_q = \gamma^q \Delta t$), such that the quantum circuit for realizing a single time step can be reused.

\begin{figure}[!ht]
\centering
\subfloat[uniform coefficients]{\includegraphics[width=0.8\columnwidth]{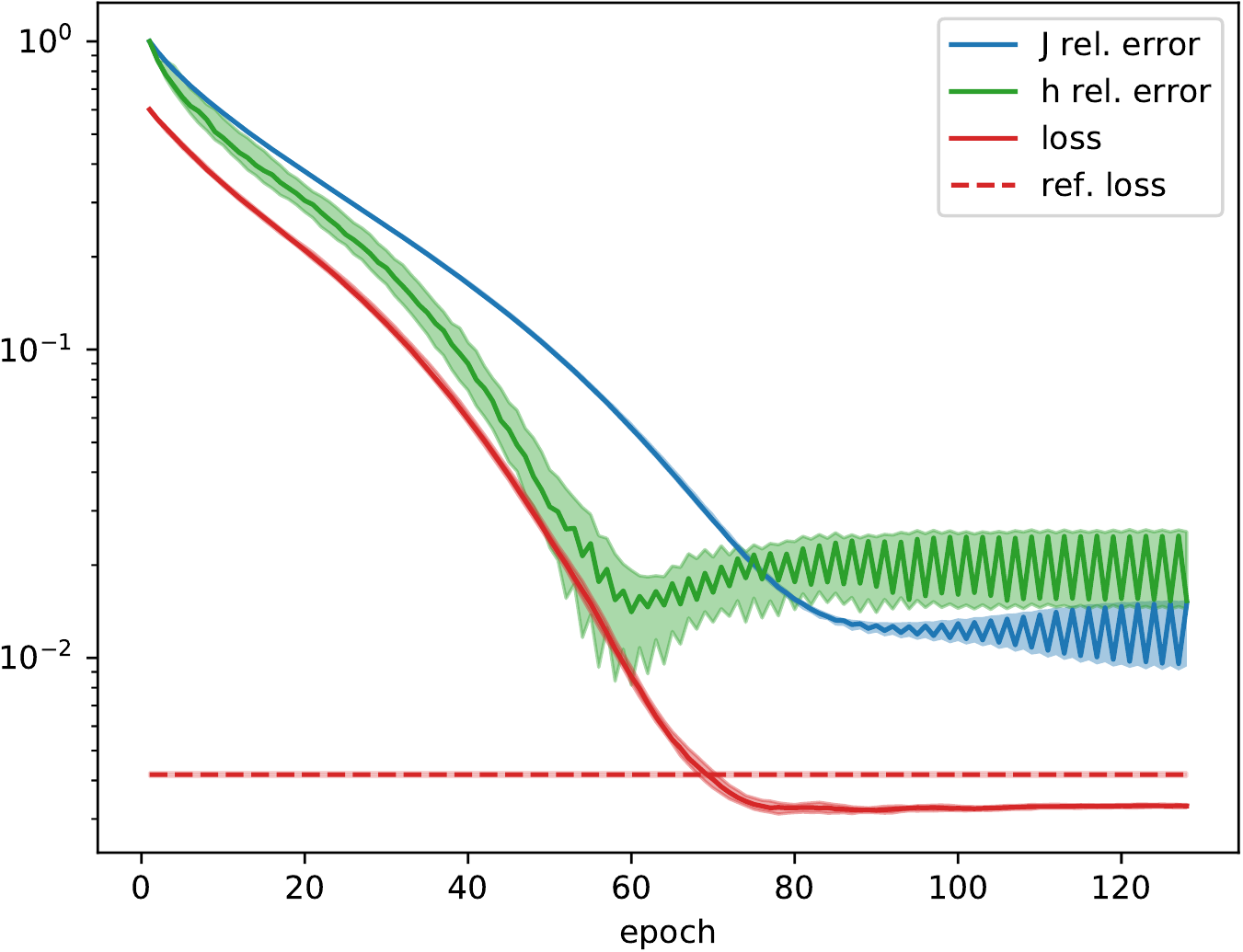}%
\label{fig:reconstruct_lattice_uniform_H_training}} \\
\subfloat[random coefficients (disorder)]{\includegraphics[width=0.8\columnwidth]{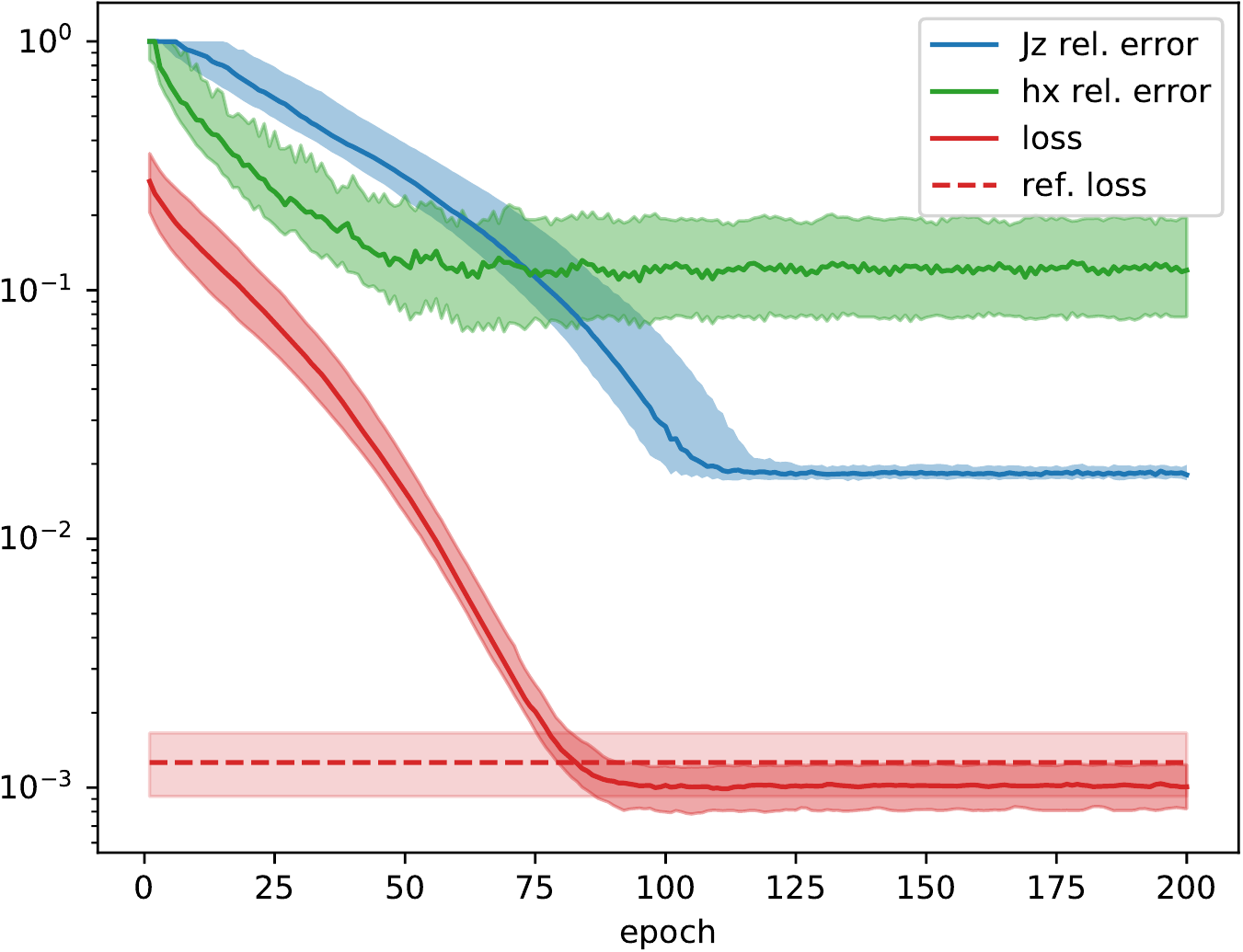}%
\label{fig:reconstruct_lattice_disordered_H_training}}
\caption{Relative reconstruction error of the Hamiltonian parameters for the $3 \times 4$ lattice model in (a) Eq.~\eqref{eq:HIsing_uniform} and (b) Eq.~\eqref{eq:HIsing_disordered}, and loss function during training, when fitting a quantum circuit representation of a time step (Fig.~\ref{fig:strang_int_local_circuit}) to the exact Born measurement averages.}
\end{figure}

We quantify the deviation between the exact Born measurement probabilities, $p(t) = \abs{\psi(t)}^2$, and the probabilities $\tilde{p}(t)$ resulting from the trotterized circuit Ansatz, via the Kullback-Leibler divergence: \begin{equation}
\label{eq:KL_meas_prob}
 D_{\text{KL}}\left(p(t) \parallel \tilde{p}(t)\right) = \sum_j p_j(t) \log \frac{p_j(t)}{\tilde{p}_j(t)}.
\end{equation}

Fig.~\ref{fig:reconstruct_lattice_uniform_H_training} visualizes the optimization progress, i.e., the relative errors of $J$ and $\vec{h}$, and the loss function \eqref{eq:KL_meas_prob}. The darker curves show medians over $100$ trials of random initial states, and the lighter shades $25\%$ quantiles. The dashed horizontal line is the loss function evaluated at the ground truth values of $J$ and $\vec{h}$; it is non-zero due to the Strang splitting approximation of a time step. Interestingly, the optimization arrives at an even smaller loss value starting around training epoch $70$. We interpret this as an artefact of the Strang splitting approximation -- note that around this epoch, the relative error of $\vec{h}$ slightly increases again. We have used the \mbox{RMSProp} optimizer \cite{rmsprop} with a learning rate of $0.005$ here. In summary, the reachable relative error is around $0.02$, and higher accuracy would likely require a smaller time step to reduce the Strang splitting error.

Next, we consider a Hamiltonian \eqref{eq:HIsing} with disorder (and external field solely in $x$-direction):
\begin{equation}
\label{eq:HIsing_disordered}
H = - \sum_{\langle j, \ell \rangle} J_{j,\ell} \, \sigma^z_j \sigma^z_\ell - \sum_{j \in \Lambda} h_j \cdot \sigma^x_j
\end{equation}
with i.i.d.\ random coefficients $J_{j,\ell}$ and $h_j$. For the numerical experiment, $J_{j,\ell}$ is uniformly distributed in the interval $[0.8, 1.2]$, and $h_j \sim 0.5\,\mathcal{N}(0, 1)$ (normal distribution).

The results are shown in Fig.~\ref{fig:reconstruct_lattice_disordered_H_training}. As before, we evaluate $100$ realizations of the experiment, with a set of coefficients and random initial state drawn independently for each trial. We take the maximum over the relative errors of $J_{j,\ell}$ for all $j, \ell$ to arrive at the relative error reported in Fig.~\ref{fig:reconstruct_lattice_disordered_H_training}, and likewise for $h_j$. The ``reference'' loss function is the KL divergence evaluated at the ground truth coefficients. It is non-zero due to Strang splitting errors, and fluctuates due to the different random coefficients at each trial.

\section{Conclusions and outlook}

Our work demonstrates the feasibility of using measurements at different time points to reconstruct the time evolution operator. The approach presented in this work has two main benefits: First, it requires only a single (or few) initial state(s) and reduces the number of different kinds of measurements for the reconstruction of the Hamiltonian. From an experimental point of view, this would be helpful when the experimental setup makes it easier to maintain the time evolution of the system than to implement a variety of measurements. Second, as demonstrated in Sect.~\ref{partial-meas}, the method allows for the reconstruction of a Hamiltonian even when only part of the system can be observed.

A natural extension of our work is QPT in the situation of dissipation and noise processes, i.e., a system governed by a Lindblad equation. This would pose the additional challenge of a limited time window for extracting information and a larger number of free parameters. An interesting alternative approach could be a mapping to a unitary evolution with an unobserved environment \cite{pokorny-2020}, which would fit into the present framework.

The relaxation method in Sect.~\ref{sec:relaxation} can be regarded as tool to smooth the optimization landscape, but an open question is how to guarantee convergence to the correct solution. This becomes particularly relevant for larger systems and the likewise increasing number of free parameters.

A related task for future work is a sensitivity analysis with respect to inaccuracies and noise in the measurement data. A modified version of Takens theorem still applies in this situation~\cite{stark-1997}, but it is not clear how accurate our algorithm can reconstruct $H$. A first step could be a gradient calculation of the measurement averages with respect to the Hamiltonian parameters.

\acknowledgments

We thank the Munich Center for Quantum Science and Technology for support.

\bibliography{qpt_delay_meas}

\end{document}